\documentclass[3p,times,authoryear]{elsarticle}

\usepackage{amsmath}
\usepackage{amssymb}
\usepackage[figuresright]{rotating}
\usepackage{url}
\usepackage{hyperref}

\hypersetup{
    colorlinks=true,
    linkcolor=black,
    citecolor=black,
    urlcolor=black
}

\makeatletter
\def\ps@pprintTitle{%
    \let\@oddhead\@empty
    \let\@evenhead\@empty
    \let\@oddfoot\@empty
    \let\@evenfoot\@empty
}
\makeatother
\begin{document}

\begin{frontmatter}

\title{Milk yield responses to heat stress in tropical dairy cattle}

\author[affil1]{Leonardo Nascimento Ferreira\corref{cor1}}
\ead{leonardo.nascimento@puc-campinas.edu.br}
\author[affil2]{Delane Ribas da Rosa}
\author[affil3]{Renata Negri}
\author[affil4]{Marcos Vinicius Gualberto Barbosa da Silva}
\author[affil5]{Rafaella Resende Andrade}
\author[affil6]{Nicole Costa Resende Ferreira}

\cortext[cor1]{Corresponding author.}

\address[affil1]{Polytechnic School, Pontifical Catholic University of Campinas, Campinas - SP, 13087-571, Brazil}
\address[affil2]{School of Veterinary and Animal Science, Federal University of Goi\'as , Goi\^ania - GO, 74690-900, Brazil}
\address[affil3]{School of Engineering, S\~ao Paulo State University, Ilha Solteira - SP, 15385-000, Brazil}
\address[affil4]{Embrapa Dairy Cattle, Juiz de Fora - MG, 36038-330, Brazil}
\address[affil5]{School of Agronomy, Federal University of Goi\'as, Goi\^ania - GO, 74690-900, Brazil}
\address[affil6]{Embrapa Territorial, Campinas - SP, 13070-115, Brazil}

\begin{abstract}
\noindent\textbf{Context:} Rising ambient temperatures and warm tropical nights increasingly threaten the productivity and resilience of dairy production systems across tropical landscapes.\\
\textbf{Objective:} This study characterises multi-scale interactions among regional thermal regimes, circadian cooling relief, endogenous metabolic heat load, and crossbred genetics on daily milk yield across Brazilian dairy systems.\\
\textbf{Methods:} A 24-year national panel (2001--2024) comprising 788,528 official test-day records from 87,809 Girolando (Holstein $\times$ Gyr) cattle across 408 municipalities was linked to meteorological monitoring. Two-way fixed effects panel models, distributed lag models, and piecewise linear threshold splines were estimated across production quartiles and certified crossbred fractions.\\
\textbf{Results and conclusions:} Although aggregate population summaries suggest stable milk yield up to a daily mean Temperature--Humidity Index (THI) of 76, production-stratified splines reveal that high-yielding animals suffer significant losses starting at the classical threshold of 68 and accelerating above 76 ($-0.162$~kg~animal$^{-1}$~d$^{-1}$~unit$^{-1}$, $p = 0.0062$), whereas low-yielding animals remain unaffected ($p = 0.81$). Distributed lag models identify an immediate test-day drop followed by a deeper trough four days later. Absence of nighttime relief ($\mathrm{THI}_{\min} \ge 68$) reduces daily yield by an additional 2.37~kg~animal$^{-1}$~d$^{-1}$ ($p \le 0.001$) and causes an 8.6\% cumulative loss after seven consecutive warm nights, whereas wide diurnal thermal amplitude buffers output by $+1.60$~kg~animal$^{-1}$~d$^{-1}$. At equivalent high yields ($\approx 31$~kg~animal$^{-1}$~d$^{-1}$), intermediate crossbreds ($1/2$ and $5/8$ Holstein) sustain production up to THI 76, whereas high-grade taurine animals ($7/8$ Holstein) decline at THI 68.\\
\textbf{Significance:} Climate adaptation in tropical dairy systems requires integrating continuous nocturnal cooling for high-yielding groups with balanced Zebu--taurine crossbreeding.
\end{abstract}

\begin{keyword}
Dairy cattle \sep Heat stress \sep Temperature--Humidity Index \sep Milk yield \sep Nighttime cooling \sep Crossbreeding \sep Distributed lag models
\end{keyword}

\end{frontmatter}

%% ===================================================================
\section{Introduction}
%% ===================================================================
Heat stress occurs when the heat load of an animal exceeds its capacity to dissipate heat to the environment. In lactating dairy animals, whose metabolic heat production rises with milk output, heat stress reduces dry matter intake, alters energy partitioning and ultimately depresses milk yield, fertility and welfare \citep{kadzere2002, west2003, collier2006, rhoads2009, polsky2017}. The associated economic losses are substantial even in temperate countries \citep{stpierre2003}, and they are expected to be larger in tropical regions where animals are exposed to high temperature and humidity for most of the year \citep{silva2002, titto2011, Ferreira2024}.

The Temperature--Humidity Index (THI) combines air temperature and relative humidity into a single indicator of thermal load and remains the most widely adopted metric for heat-stress assessment in dairy cattle \citep{nrc1971, ravagnolo2000, bohmanova2007}. Classical guidelines placed the onset of heat stress at a THI of about 72 \citep{armstrong1994}, but subsequent work with high-producing animals indicated that yield losses may start at THI values as low as 68 \citep{zimbelman2009}. Threshold values derived from controlled or regional studies, however, do not necessarily transfer to commercial dairy systems that differ in genetic merit, housing, cooling infrastructure and regional climate \citep{bohmanova2007, bernabucci2014}.

Brazil is one of the largest milk producers worldwide, and its dairy production systems are distributed across climatic landscapes ranging from humid subtropical in the South to tropical savannah (Cerrado) and semi-arid climates in the Centre-West and North-East \citep{Ferreira2024, FERREIRA2026108392}. Official milk-recording programmes generate large databases of test-day records that, when matched with meteorological information, offer a unique opportunity to evaluate multi-scale interactions between environmental exposure and livestock performance across whole-farm and regional levels. Crossbreeding between Holstein (\emph{Bos taurus}) and Gyr (\emph{Bos indicus}) has been the cornerstone of tropical dairy systems in Brazil, giving rise to the Girolando composite breed that combines high milk yield potential with tropical adaptation \citep{santana2017, silva2021}. Observational field data from commercial systems, however, are affected by confounding factors that are absent in controlled trials: the lactation curve \citep{wood1967}, seasonal calving and forage patterns, long-term genetic and management intensification trends, and the uneven geographical distribution of herds.

From an agricultural systems perspective, thermal resilience in tropical dairying is governed by interactions across hierarchical levels: individual physiological state and endogenous heat production, genetic composition and retained heterosis, farm-level cooling and lactation management, and landscape-level circadian thermal regimes. The central hypothesis of this study is that the apparent high thermal threshold observed in aggregate tropical field data masks substantial heterogeneity across these system components. Specifically, it is hypothesised that high-yielding animals experience production losses at classical temperate thresholds ($\mathrm{THI} \approx 68$), that the absence of nighttime thermal relief compounds multi-day yield depression, and that intermediate Holstein $\times$ Gyr crossbreds retain superior thermal buffering compared with high-grade taurine animals operating at equivalent production levels.

To test these hypotheses and advance the quantitative analysis of climate resilience in tropical livestock systems, the objective of this study is to provide a comprehensive empirical and econometric characterisation of the relationship between daily THI and milk yield across Brazilian dairy herds. Specifically, this work: (i) mapped the spatial distribution of recorded production and of annual heat-stress exposure across municipalities; (ii) described the temporal and seasonal dynamics of THI and yield over 24 years; (iii) examined yield across thermal zones within lactation stages; (iv) characterised the continuous THI--yield response curve to identify empirical inflection points; (v) evaluated lagged and cumulative exposure to heat stress through distributed lag models, rolling windows and multi-day persistence; (vi) quantified the protective role of diurnal thermal amplitude and nighttime cooling relief, evaluating the cumulative impact of multi-night tropical persistence; (vii) evaluated production-level heterogeneity across 305-day cumulative milk yield quartiles, testing for empirical threshold shifts through piecewise linear splines; and (viii) evaluated breed composition and retained heterosis across Holstein $\times$ Gyr crossbreds ($1/4$ to $7/8$ Holstein fraction), implementing a high-yield decoupling stress test to isolate genetics from baseline productivity.

The remainder of this paper is structured as follows. Section~\ref{sec:methods} describes the milk-recording and meteorological datasets, the bioclimatic indicators and thermal categories, and the econometric and statistical modelling strategies. Section~\ref{sec:results} presents the empirical results, including spatial maps, seasonal patterns, continuous response functions, distributed lag dynamics, nocturnal relief effects, and stratifications by production potential and crossbred composition. Section~\ref{sec:discussion} discusses the biophysical and physiological mechanisms, evaluates the systems-level implications for farm management, cooling infrastructure and breeding, and outlines study limitations. Finally, Section~\ref{sec:conclusions} summarises the main conclusions and practical recommendations.

%% ===================================================================
\section{Material and methods}\label{sec:methods}
%% ===================================================================

\subsection{Data}
The dataset comprised official test-day records of individual animals collected between January 2001 and December 2024 from the official milk-recording service (\textit{Controle Leiteiro Oficial}) of the Brazilian Association of Girolando Breeders. Because milk production is a sex-limited physiological trait, all records pertained exclusively to biologically female lactating dairy cattle. Each record contained the unique animal identifier, test date, daily milk yield (kg~animal$^{-1}$ d$^{-1}$), days in milk (DIM), number of daily milkings, genealogical breed composition, and the municipality of the herd according to the Brazilian Institute of Geography and Statistics (IBGE) code. Quality control procedures excluded records with missing production or meteorological observations, lactations outside the standard 5--305 DIM interval, biologically implausible daily yields ($<0.1$ or $>80$~kg~animal$^{-1}$ d$^{-1}$), and herds lacking verified geographic coordinates. The resulting dataset comprised 788,528 test-day records ($N = 788{,}528$) from 87,809 animals located in 408 municipalities across 19 Brazilian states and the Federal District, spanning 7,508 distinct test dates. The records were primarily concentrated in the principal dairy basins of Minas Gerais (64.8\%), Goi\'as (17.7\%) and S\~ao Paulo (8.2\%). Each animal contributed an average of 9.0 test-day records (median 8), consistent with monthly recording across a standard lactation cycle. All records fell within 5--305 DIM, with 83.0\% of records obtained from animals milked twice daily and 17.0\% milked three times daily.

\subsection{Temperature--Humidity Index and thermal zones}
Meteorological data were obtained from the National Institute of Meteorology (INMET, \textit{Instituto Nacional de Meteorologia}), which maintains an extensive surface network of automatic and conventional weather stations across Brazil. Daily records of dry-bulb air temperature ($T_{\mathrm{db}}$, $^\circ$C) and relative humidity ($\mathrm{RH}$, \%) were extracted and spatially matched to the municipality centroid of each dairy herd. The THI was calculated according to the standard National Research Council formulation \citep{nrc1971}, as given in Eq.~(\ref{eq:thi}):
\begin{equation}\label{eq:thi}
\mathrm{THI} = 0.8 \, T_{\mathrm{db}} + (\mathrm{RH} / 100)(T_{\mathrm{db}} - 14.4) + 46.4.
\end{equation}
For each municipality and date, daily mean THI ($\mathrm{THI}_{\mathrm{mean}}$), daily maximum THI ($\mathrm{THI}_{\max}$), and diurnal thermal amplitude ($\Delta\mathrm{THI}$) were computed. Records were classified into five thermal zones adapted from \citet{armstrong1994} and \citet{zimbelman2009}: Comfort ($\mathrm{THI} \le 68$), Alert ($68 < \mathrm{THI} \le 72$), Mild ($72 < \mathrm{THI} \le 76$), Moderate ($76 < \mathrm{THI} \le 80$) and Severe ($\mathrm{THI} > 80$). For lactation-stratified analyses, records were grouped into three canonical stages: Peak (0--100~d), Mid (101--200~d), and Late (201--305~d).

\subsection{Spatial analysis}
Two municipal indicators were mapped on the 2024 IBGE municipal boundaries. The \emph{mean annual cumulative recorded yield} (t~year$^{-1}$) was obtained by summing all test-day yields recorded in a municipality within each calendar year and averaging these annual totals across the years in which the municipality was monitored. Because it accumulates test-day records, this indicator reflects both herd productivity and recording intensity, and should be interpreted as an index of recorded production rather than as total municipal milk output. The \emph{mean annual number of heat-stress days} was computed from unique municipality--test-date pairs (to avoid weighting by the number of animals tested on a given day) as the proportion of dates with daily mean $\mathrm{THI} \ge 72$ (Mild zone or above) in each municipality--year, scaled to an annual basis (proportion $\times$ 365.25 days) and averaged across years.

\subsection{Temporal, stratified and dose--response analyses}
Monthly averages of daily mean THI, daily maximum THI and milk yield were computed for each calendar month of the study period (long-term trajectory), and monthly averages of daily mean THI and yield were pooled across years to characterise the seasonal profile, reported with $\pm 1$ standard deviation (SD). The overall and annual distributions of thermal zones were expressed as the percentage of test-day records in each zone. Median yield was tabulated for each combination of lactation stage and thermal zone.

The continuous response curve was obtained by grouping records into 0.5-unit THI bins and computing the median, mean and interquartile range (IQR) of yield within each bin; bins with fewer than 100 records ($n < 100$) and those exceeding THI 80.5 were discarded to eliminate small-sample compositional artefacts from extreme outliers. The cumulative yield decline at each THI bin above 68 was defined as the difference between the median yield of the first bin at or above THI 68 (reference, 20.8~kg~animal$^{-1}$ d$^{-1}$) and the median yield of the bin.

\subsection{Lagged and cumulative heat stress modelling}
To investigate the temporal dynamics of thermal stress, daily THI values for up to six days preceding the test day ($t_{-1}, \dots, t_{-6}$) were reconstructed analytically from the rolling mean THI series ($\mathrm{THI}_{k\mathrm{d}}$) available in the dataset according to Eq.~(\ref{eq:lag}):
\begin{equation}\label{eq:lag}
\mathrm{lag}_k = (k+1) \, \mathrm{THI}_{(k+1)\mathrm{d}} - k \, \mathrm{THI}_{k\mathrm{d}}, \quad \forall k \in \{1, \dots, 6\},
\end{equation}
where $\mathrm{lag}_0 = \mathrm{THI}_{t_0} = \mathrm{THI}_{1\mathrm{d}}$ represents the test-day mean THI at lag order $k = 0$.

To isolate the marginal contribution of each preceding day without omitted-variable bias, a distributed lag model (DLM) with two-way fixed effects (TWFE) was estimated \citep{gasparrini2010}, as defined in Eq.~(\ref{eq:dlm}):
\begin{equation}\label{eq:dlm}
y_{it} = \alpha_i + \gamma_{\mathrm{month}(t)} + \sum_{k=0}^{6} \beta_k \, \mathrm{lag}_{k, it} + \delta_1 (\mathrm{DIM}_{it} / 100) + \delta_2 (\mathrm{DIM}_{it} / 100)^2 + \epsilon_{it},
\end{equation}
where $y_{it}$ is the daily milk yield of animal $i$ at test date $t$, $\alpha_i$ is the individual animal fixed effect, $\gamma_{\mathrm{month}(t)}$ denotes calendar-month indicator dummies absorbing macro-seasonal calving and forage cycles, $\beta_k$ is the marginal regression slope at lag $k$, $\delta_1$ and $\delta_2$ are polynomial regression coefficients for DIM modelled as a second-order polynomial, and $\epsilon_{it}$ is the idiosyncratic error term. Because official milk recording is conducted at monthly intervals, this specification relates each point-in-time test-day yield ($t_0$) to the daily meteorological history of the six days immediately preceding that test ($t_{-1}$ to $t_{-6}$), rather than tracking consecutive daily milk weights of the same individual. For comparison, a model with animal fixed effects (FE) alone (unadjusted for calendar month) was also fitted. The long-run cumulative effect was computed as $\sum_{k=0}^{6} \beta_k$.

Additionally, the analysis evaluated: (i) cumulative rolling windows of $1$ to $7$ days for both mean and maximum daily THI; (ii) heat-stress persistence, defined as the count of days within the preceding seven days in which daily THI reached or exceeded moderate ($\ge 72$) and severe ($\ge 76$) thresholds; and (iii) heterogeneity of the lag-response profile across lactation stages: Peak (0--100~d), Mid (101--200~d) and Late (201--305~d).

\subsection{Diurnal thermal amplitude and nighttime relief modelling}
In addition to daily mean and rolling exposure, circadian thermal fluctuations dictate whether animals can dissipate daytime heat loads during the night. The minimum daily THI ($\mathrm{THI}_{\min}$) for each municipality and date was reconstructed from daily maximum THI ($\mathrm{THI}_{\max}$) and diurnal thermal amplitude ($\Delta\mathrm{THI}$) via Eq.~(\ref{eq:min_thi}):
\begin{equation}\label{eq:min_thi}
\mathrm{THI}_{\min} = \mathrm{THI}_{\max} - \Delta\mathrm{THI}.
\end{equation}
Following physiological criteria for nocturnal heat dissipation \citep{scott1983, igono1992, honig2012}, records were classified into four nighttime relief categories: Full relief ($\mathrm{THI}_{\min} < 64$), Partial relief ($64 \le \mathrm{THI}_{\min} < 68$), Mild tropical night ($68 \le \mathrm{THI}_{\min} < 72$), and Severe tropical night ($\mathrm{THI}_{\min} \ge 72$). Dates with $\mathrm{THI}_{\min} \ge 68$ were designated as ``tropical nights'', during which ambient conditions remain above the thermoneutral threshold continuously for 24 hours. Diurnal thermal amplitude was categorised into terciles: Low ($\Delta\mathrm{THI} < 10.5$), Medium ($10.5 \le \Delta\mathrm{THI} < 14.5$), and High ($\Delta\mathrm{THI} \ge 14.5$).

To evaluate the non-linear interaction between daytime heat and nighttime recovery, median milk yield was examined across a two-dimensional grid of daytime heat ($\mathrm{THI}_{\max} \in \{\le 72, (72, 76], (76, 80], > 80\}$) and nocturnal relief. The protective buffering of diurnal amplitude was also assessed across continuous daily mean THI (1.0-unit bins from 64 to 80). Furthermore, multi-night tropical persistence was defined as the cumulative count of tropical nights ($\mathrm{THI}_{\min} \ge 68$) over the preceding seven days ($0$ to $7$ nights), evaluated overall and stratified by lactation stage: Peak (0--100~d), Mid (101--200~d), and Late (201--305~d).

Finally, to estimate the isolated marginal effect of lacking nighttime relief while controlling for daytime peak heat, individual animal genetics, and seasonal cycles, a TWFE model was specified according to Eq.~(\ref{eq:night}):
\begin{equation}\label{eq:night}
y_{it} = \alpha_i + \gamma_{\mathrm{month}(t)} + \beta_1 \, \mathrm{THI}_{\max, it} + \beta_2 \, \mathbb{I}(\mathrm{THI}_{\min, it} \ge 68) + \delta_1 (\mathrm{DIM}_{it} / 100) + \delta_2 (\mathrm{DIM}_{it} / 100)^2 + \epsilon_{it},
\end{equation}
where $\mathbb{I}(\mathrm{THI}_{\min, it} \ge 68)$ is an indicator function for the complete absence of nocturnal thermal relief, and $\beta_1$ and $\beta_2$ denote the marginal effects of daytime maximum THI and tropical nights, respectively.

\subsection{Production-level heterogeneity and threshold spline modelling}
To determine whether endogenous metabolic heat production ($H_{\mathrm{met}}$) modulates thermal sensitivity and accounts for the high inflection threshold observed in field data, records were stratified by individual mature-equivalent 305-day milk production potential ($\mathrm{PROD\_305}$). Animals were partitioned into population quartiles ($Q_1\text{--}Q_4$): $Q_1$ (Low, $<4,389$~kg, mean daily yield 13.7~kg~animal$^{-1}$ d$^{-1}$, $N = 197{,}140$), $Q_2$ (Mid-Low, $4,389\text{--}5,847$~kg, mean 18.5~kg~animal$^{-1}$ d$^{-1}$, $N = 197{,}127$), $Q_3$ (Mid-High, $5,847\text{--}7,494$~kg, mean 22.9~kg~animal$^{-1}$ d$^{-1}$, $N = 197{,}133$), and $Q_4$ (High, $>7,494$~kg, mean 30.8~kg~animal$^{-1}$ d$^{-1}$, $N = 197{,}128$).

For each quartile, TWFE models controlling for individual animal characteristics and macro-seasonal calendar month-by-year shocks were estimated using Eq.~(\ref{eq:twfe}):
\begin{equation}\label{eq:twfe}
y_{it} = \alpha_i + \gamma_{\mathrm{period}(t)} + \beta \, \mathrm{THI}_{it} + \sum_{m=1}^{3} \delta_m (\mathrm{DIM}_{it} / 100)^m + \epsilon_{it},
\end{equation}
where $\gamma_{\mathrm{period}(t)}$ represents unique calendar month-by-year fixed effects, $\beta$ is the marginal THI slope, and DIM is modelled as a third-order polynomial ($m \in \{1, 2, 3\}$) with coefficients $\delta_m$ to capture the lactation trajectory with high precision. Models were estimated for immediate daily THI (1-day) and cumulative rolling THI (7-day).

To formally identify the critical breakpoint where production losses commence across production tiers, piecewise linear threshold splines were estimated as shown in Eq.~(\ref{eq:spline}):
\begin{equation}\label{eq:spline}
y_{it} = \alpha_i + \gamma_{\mathrm{period}(t)} + \beta_T \max(0, \mathrm{THI}_{it} - T) + \sum_{m=1}^{3} \delta_m (\mathrm{DIM}_{it} / 100)^m + \epsilon_{it},
\end{equation}
for candidate thresholds $T \in \{68, 70, 72, 74, 76\}$. The parameter $\beta_T$ measures the marginal slope (kg~animal$^{-1}$ d$^{-1}$~unit$^{-1}$) beyond threshold $T$. Expanding rolling windows (1, 2, 3, 5, and 7 days) were also estimated to evaluate the trajectory of cumulative physiological exhaustion.

\subsection{Breed composition, crossbred heterosis, and decoupling stress test}
To evaluate the biophysical protection conferred by Zebu (\emph{Bos indicus}) genetics and heterosis, animals were classified into six certified genealogical crossbred groups based on their official proportion of Holstein (\emph{Bos taurus}, Hol) and Gyr (\emph{Bos indicus}) inheritance: $1/4$~Hol (75\% Zebu, $N = 3{,}091$), $3/8$~Hol (62.5\% Zebu, $N = 8{,}632$), $1/2$ first filial generation crossbred (F1, 50\% Zebu, 100\% retained heterosis, $N = 39{,}410$), $5/8$~Girolando (37.5\% Zebu, stabilised composite, $N = 373{,}737$), $3/4$~Hol (25\% Zebu, $N = 100{,}283$), and $7/8$~Hol (12.5\% Zebu, $N = 78{,}574$). These six certified crossbred categories accounted for 603,727 test-day records (76.6\% of the total dataset); the remaining 184,801 records comprised purebred animals (Holstein or Gyr) and non-standard crossbred fractions, which were retained in aggregate population models but excluded from breed-specific heterosis analyses.

Because higher Holstein fractions are typically associated with higher baseline milk production (which inherently elevates endogenous metabolic heat generation), direct cross-sectional comparisons between breeds risk confounding genetic thermal tolerance with production scale \citep{collier2006, santana2017}. To disentangle these two mechanisms, a high-yield decoupling stress test was designed by isolating the subpopulation of animals in the top production quartile ($Q_4$: $\mathrm{PROD\_305} \ge 7,494$~kg). Within this stratum, animals across all crossbred categories produce at virtually identical high daily yields ($\approx 31$~kg~animal$^{-1}$ d$^{-1}$), ensuring that metabolic heat output is held constant while genetic composition and heterosis vary. Within both the general population and the decoupled $Q_4$ subpopulation, TWFE models (Eq.~(\ref{eq:twfe})) and piecewise linear threshold splines (Eq.~(\ref{eq:spline})) were estimated across candidate breakpoints ($T \in \{68, 70, 72, 74, 76\}$).

\subsection{Statistical analysis and computational implementation}
Descriptive associations were evaluated using Spearman's rank correlation ($\rho$) and Pearson's linear correlation ($r$). Yield distributions across thermal zones and categories were compared using two-sided independent-samples Student's $t$-tests (reported via the test statistic $t_{\mathrm{stat}}$ and probability value $p$) with Welch's correction when variances were unequal, and standardised effect sizes were reported as Cohen's $d$. Point estimates in regression models were reported with 95\% confidence intervals (CI). Significance levels were denoted as ns ($p > 0.05$), $^{*}$ ($p \le 0.05$), $^{**}$ ($p \le 0.01$), $^{***}$ ($p \le 0.001$) and $^{****}$ ($p \le 10^{-4}$).

Econometric panel specifications with individual animal fixed effects ($\alpha_i$) and calendar month indicators were estimated via within-animal demeaning, while specifications incorporating high-dimensional animal and month-by-year period fixed effects ($\gamma_{\mathrm{period}(t)}$) were estimated using the method of alternating projections (Gauss--Seidel iterative demeaning) \citep{gaure2013} with standard errors adjusted using the cluster-robust variance estimator (CR1) clustered at the individual animal level ($G = 87{,}809$ clusters) \citep{wooldridge2010}.

%% ===================================================================
\section{Results}\label{sec:results}
%% ===================================================================

\subsection{Dataset overview}
The primary descriptive characteristics of the monitored herds, including the 5th and 95th percentiles (P5 and P95), are summarised in Table~\ref{tab:desc}. Daily milk production exhibited wide dispersion around an overall average of approximately 21.5~kg~animal$^{-1}$ d$^{-1}$, reflecting the co-existence of diverse management intensities and crossbred genotypes across the country. Ambient thermal conditions averaged near the upper boundary of the classical comfort zone (daily mean $\mathrm{THI} \approx 70$), while DIM values were approximately uniformly distributed across the standard lactation cycle.

\begin{table}[htbp]
\centering
\caption{Descriptive statistics of the main variables ($N = 788{,}528$ test-day records; 87,809 animals).}
\label{tab:desc}
\begin{tabular}{lrrrrr}
\hline
\textbf{Variable} & \textbf{Mean} & \textbf{Median} & \textbf{SD} & \textbf{P5} & \textbf{P95} \\
\hline
Daily milk yield (kg~animal$^{-1}$ d$^{-1}$) & 21.46 & 20.60 & 8.82 & 8.4 & 37.5 \\
Daily mean THI                               & 69.68 & 70.39 & 4.42 & 61.6 & 76.0 \\
DIM (d)                                      & 143.5 & 139.0 & 82.2 & 20 & 281 \\
\hline
\multicolumn{6}{l}{\footnotesize Note: THI, Temperature--Humidity Index; DIM, days in milk; SD, standard deviation; P5 and P95, 5th and 95th percentiles.}
\end{tabular}
\end{table}

\subsection{Spatial distribution of production and heat-stress exposure}
\begin{figure}[htbp]
    \centering
    \includegraphics[width=0.75\linewidth]{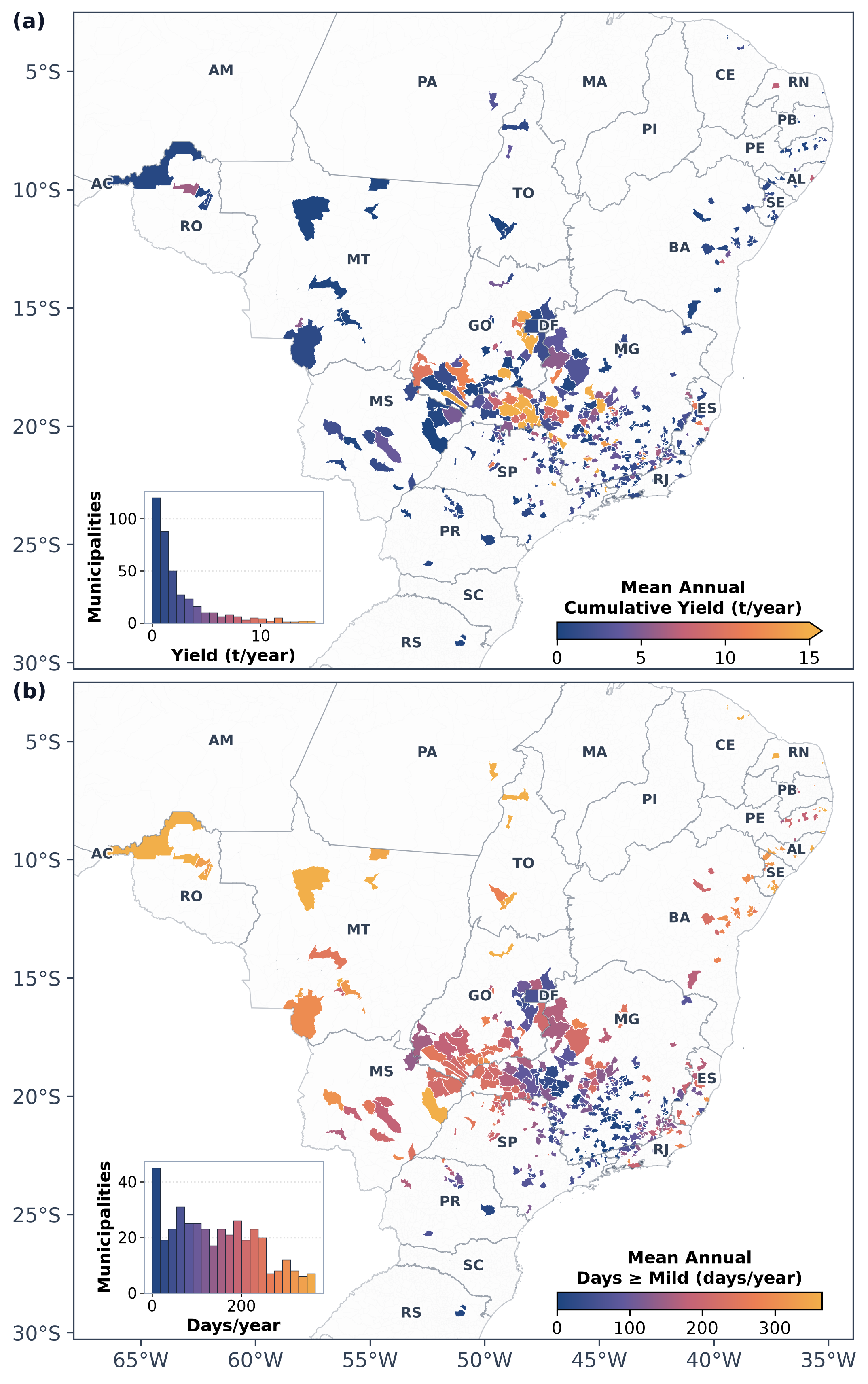}
    \caption{Spatial distribution of the 408 monitored municipalities. (a) Mean annual cumulative recorded milk yield (t~year$^{-1}$; sum of test-day yields per municipality and year, averaged across monitored years; colour scale capped at 15~t~year$^{-1}$). (b) Mean annual number of days with daily mean $\mathrm{THI} \ge 72$ (Mild zone or above), computed from unique municipality--test-date pairs and scaled to an annual basis. Insets show the distribution across municipalities and the median (dashed line). Map lines delineate study areas and do not necessarily depict accepted national boundaries.}
    \label{fig:maps}
\end{figure}

Milk production and heat-stress exposure showed clear regional patterns across Brazil (Fig.~\ref{fig:maps}).
Recorded production was heavily concentrated in the primary dairy basins of Minas Gerais, southern Goi\'as, and eastern S\~ao Paulo (Fig.~\ref{fig:maps}a). In contrast, annual exposure to heat-stress conditions exhibited a pronounced latitudinal and altitudinal gradient (Fig.~\ref{fig:maps}b). Herds situated in the equatorial North, North-East, and Centre-West lowlands were exposed to elevated thermal indices on the majority of days throughout the year. Conversely, highland dairy regions in the South-East and South enjoyed substantial bioclimatic relief, experiencing heat stress on less than half the year. Overall, municipal milk recording volume and thermal exposure were only weakly associated, confirming that Brazil's principal recorded dairy herds are concentrated in temperate upland plateaus rather than the hottest tropical lowlands.

\subsection{Temporal and seasonal dynamics}
\begin{figure}[htbp]
    \centering
    \includegraphics[width=0.92\linewidth]{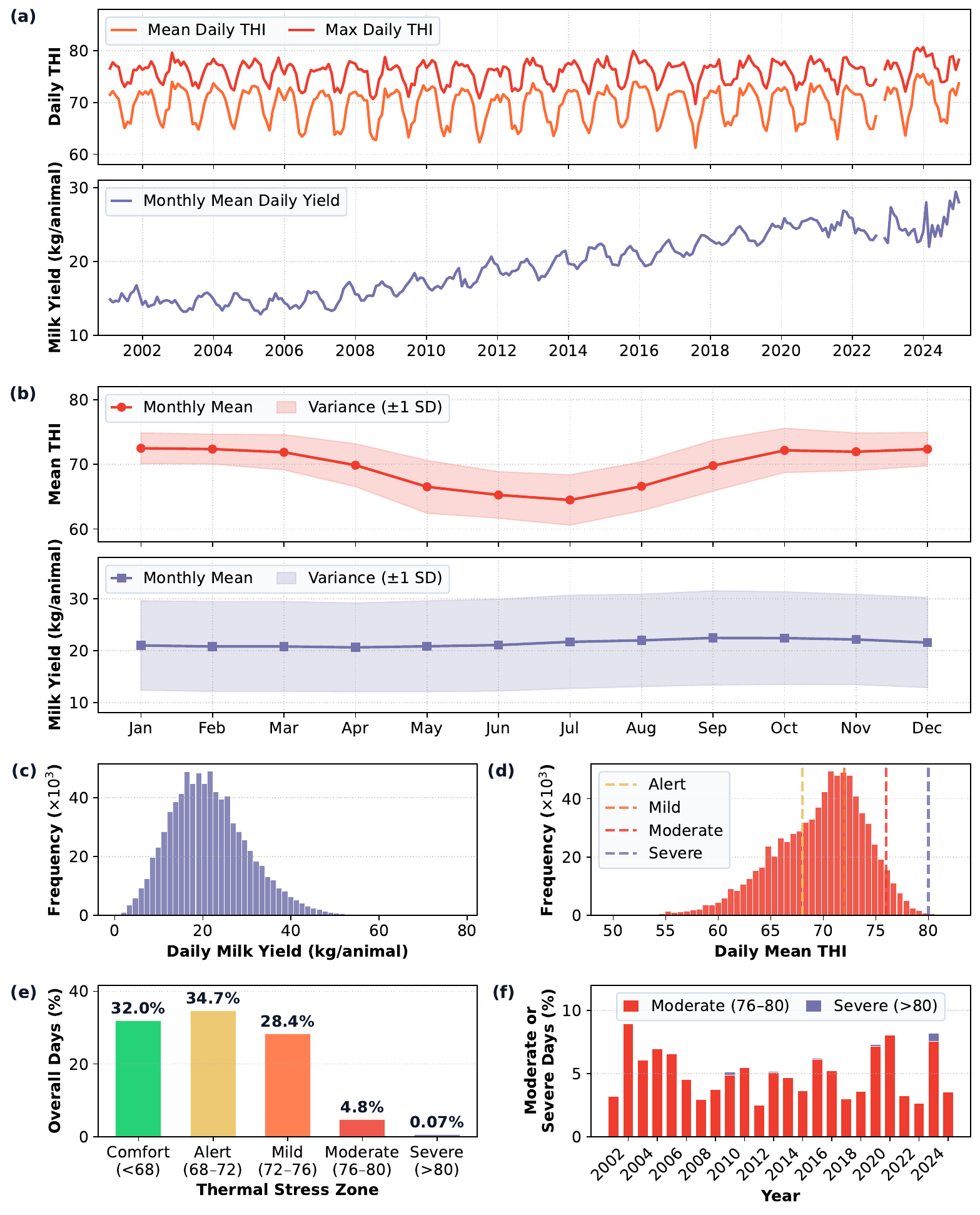}
    \caption{Descriptive overview of the dataset. (a) Monthly mean of daily mean and maximum THI (top) and monthly mean daily milk yield (bottom), 2001--2024. (b) Seasonal profile of daily mean THI (top) and daily milk yield (bottom) by month of the year: mean (line) and $\pm 1$ SD (shaded area). (c) Frequency distribution of daily milk yield (mean: 21.5~kg~animal$^{-1}$ d$^{-1}$; median: 20.6~kg~animal$^{-1}$ d$^{-1}$). (d) Frequency distribution of daily mean THI (mean: 69.7; median: 70.4) with thermal zone thresholds. (e) Percentage of test-day records in each thermal zone. (f) Annual percentage of days with Moderate (THI 76--80) and Severe ($\mathrm{THI} > 80$) heat stress.}
    \label{fig:descriptive}
\end{figure}

The multi-decadal and seasonal dynamics of environmental conditions and herd productivity are illustrated in Fig.~\ref{fig:descriptive}.
Over the 24-year monitoring period, monthly mean and maximum daily THI exhibited regular annual cycles between comfortable winter troughs and summer peaks (Fig.~\ref{fig:descriptive}a).
Despite short-term meteorological fluctuations, no systematic long-term warming trend was detected in regional mean THI.
In sharp contrast, daily milk yield showed a continuous and substantial upward trajectory across the study period ($\approx 0.56$~kg~animal$^{-1}$ d$^{-1}$ per year), rising from approximately 15~kg~animal$^{-1}$ d$^{-1}$ in the early 2000s to over 25~kg~animal$^{-1}$ d$^{-1}$ by 2024.
This secular trend reflects cumulative genetic improvement, the progressive adoption of confinement housing systems (such as compost-bedded pack barns and free-stalls), and widespread enhancements in herd management and nutritional strategies.

Seasonal profiles (Fig.~\ref{fig:descriptive}b) revealed an asynchronous relationship between thermal load and milk output. While THI reached its annual peak during the rainy summer months and its minimum in mid-winter, daily milk yield peaked in late winter and early spring rather than during the period of lowest thermal load. This counter-intuitive pattern is driven by reproductive seasonality: concentrated autumn and winter calving resulted in a high proportion of animals reaching peak lactation during late winter, temporarily masking warm-season thermal suppression in unadjusted monthly means. Across the entire historical series, the vast majority of test-day records occurred under comfort, alert, or mild thermal conditions, whereas moderate and severe stress were relatively rare, emerging primarily during historic summer heatwaves such as 2019 and 2023 (Fig.~\ref{fig:descriptive}c--f).

\subsection{Yield across thermal zones and lactation stages}
Categorising records into bioclimatic zones (Table~\ref{tab:zones}) demonstrated that unadjusted daily milk output remained largely unchanged across comfort, alert, and mild conditions, showing noticeable reductions only when ambient conditions escalated to the moderate and severe stress zones.

\begin{table}[htbp]
\centering
\small
\caption{Records and daily milk yield by thermal zone (daily mean THI).}
\label{tab:zones}
\begin{tabular}{lrrrrcrrr}
\hline
\textbf{Zone}  & \textbf{$N$ records} & \textbf{\%} & \textbf{Yield Mean} & \textbf{Median} & \textbf{IQR} & \textbf{THI Mean} & \textbf{Animals} & \textbf{Municipalities} \\
\textbf{(THI)} &                    &             & \textbf{(kg~d$^{-1}$)} & \textbf{(kg~d$^{-1}$)} & \textbf{(kg~d$^{-1}$)} & \textbf{(unit)}   &               &                         \\
\hline
Comfort ($\le 68$) & 252,129 & 32.0 & 21.43 & 20.60 & 15.0--27.0 & 64.5 & 74,548 & 349 \\
Alert (68--72)     & 273,800 & 34.7 & 21.68 & 20.86 & 15.2--27.2 & 70.2 & 79,651 & 379 \\
Mild (72--76)      & 223,791 & 28.4 & 21.42 & 20.60 & 15.1--26.8 & 73.6 & 72,921 & 382 \\
Moderate (76--80)  &  38,224 &  4.8 & 20.41 & 19.20 & 14.1--25.6 & 77.1 & 18,268 & 206 \\
Severe ($>80$)     &     584 & 0.07 & 20.04 & 18.60 & 13.6--24.6 & 80.5 &    489 &  17 \\
\hline
\multicolumn{9}{l}{\footnotesize Note: THI, Temperature--Humidity Index; IQR, interquartile range (25th--75th percentiles).}
\end{tabular}
\end{table}

\begin{figure}[htbp]
    \centering
    \includegraphics[width=0.85\linewidth]{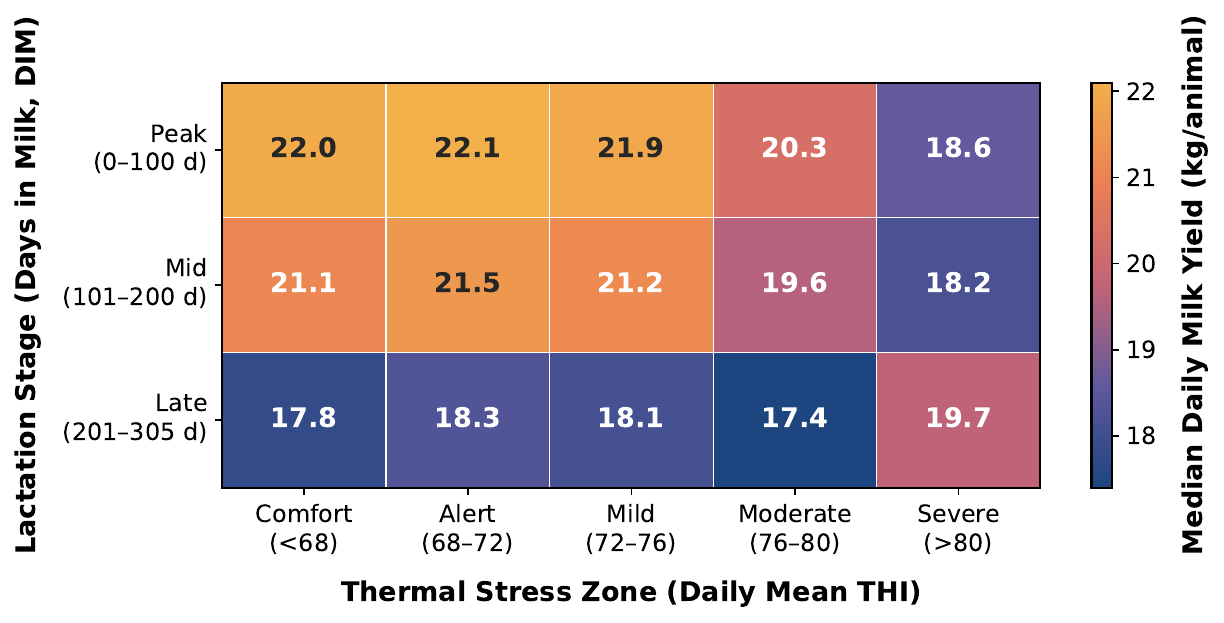}
    \caption{Median daily milk yield (kg~animal$^{-1}$ d$^{-1}$) by lactation stage (days in milk, DIM) and thermal zone (daily mean THI).}
    \label{fig:heatmap}
\end{figure}

Stratifying by lactation stage isolated the confounding influence of the lactation curve (Fig.~\ref{fig:heatmap}). In peak (0--100~d) and mid-lactation (101--200~d), median yield remained elevated and stable across comfort, alert, and mild zones, but dropped steeply under moderate and severe heat, suffering declines of 15.5\% (from 22.0 to 18.6~kg~animal$^{-1}$ d$^{-1}$) and 13.7\% (from 21.1 to 18.2~kg~animal$^{-1}$ d$^{-1}$), respectively. Conversely, animals in late lactation (201--305~d) exhibited minimal thermal sensitivity across comfort through moderate conditions (17.8 to 17.4~kg~animal$^{-1}$ d$^{-1}$), consistent with their lower metabolic heat load and diminished secretory drive. Because the severe stress zone ($\mathrm{THI} > 80$) comprises only 584 records ($0.07\%$ of the dataset across 17 municipalities), statistical power in this extreme tail is restricted; in particular, the slight yield elevation observed in late lactation under severe heat (19.7~kg~animal$^{-1}$ d$^{-1}$, $N = 213$) reflects compositional selection bias---specifically, the retention of higher-yielding animals on extended milking protocols in intensive herds---rather than genuine physiological tolerance.

\subsection{Continuous response to THI}
\begin{figure}[htbp]
    \centering
    \includegraphics[width=0.98\linewidth]{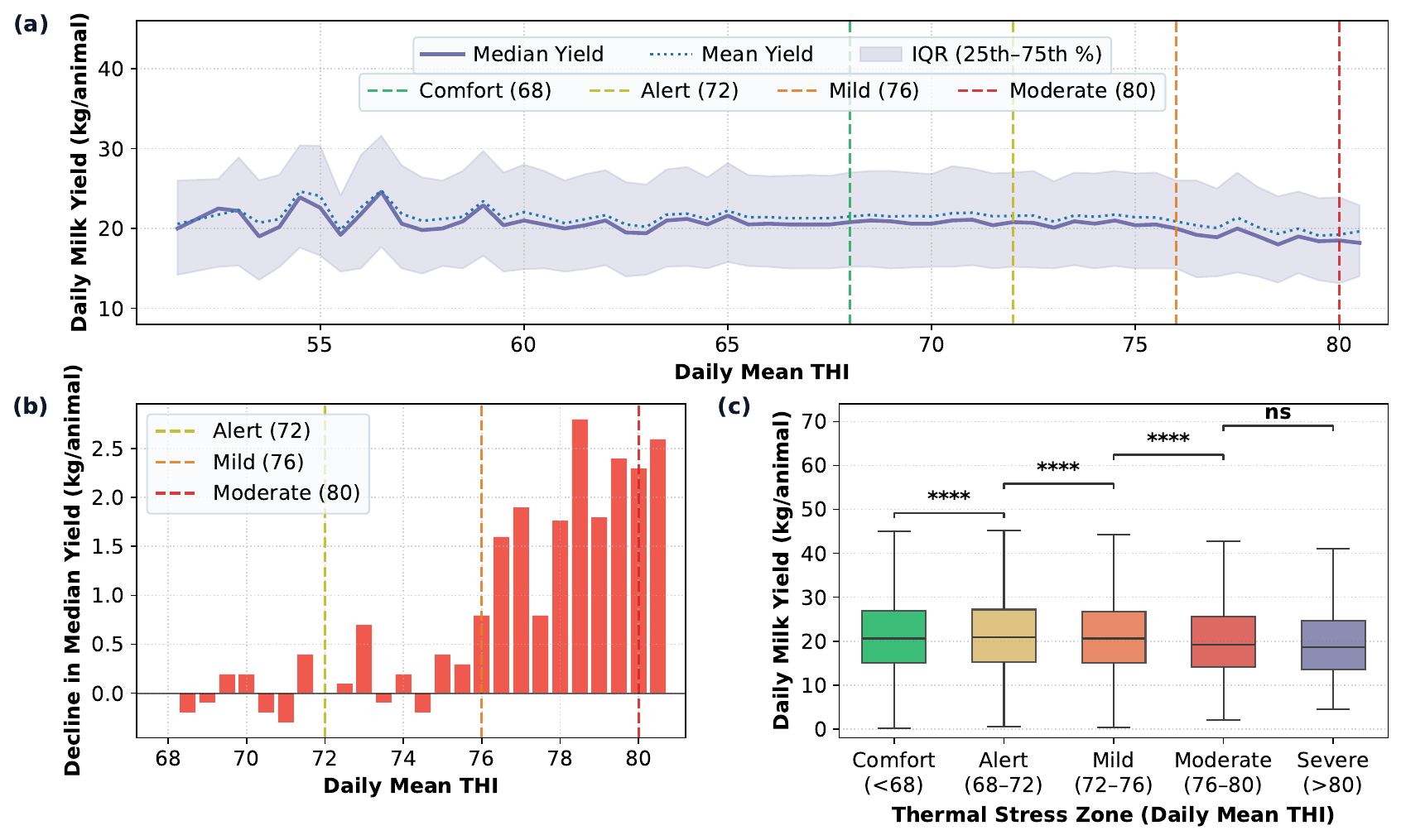}
    \caption{Continuous association between daily mean THI and daily milk yield. (a) Median (solid line), mean (dotted line) and interquartile range (IQR, shaded area) of daily milk yield in 0.5-unit THI bins ($n \ge 100$ records per bin); vertical dashed lines mark the thresholds 68, 72, 76 and 80. (b) Decline in median yield relative to the median at THI 68 (positive values indicate lower yield). (c) Distribution of daily milk yield by thermal zone (box plots, outliers omitted), with Student's $t$-tests between adjacent zones (ns: $p > 0.05$; $^{****}$: $p \le 10^{-4}$).}
    \label{fig:response}
\end{figure}

The continuous dose--response curve (Fig.~\ref{fig:response}a) confirmed a biphasic relationship between environmental thermal load and herd milk yield. Throughout the broad range from low temperatures up to mild conditions, median milk yield hovered stably around 20--21~kg~animal$^{-1}$ d$^{-1}$ without evidence of thermal impairment. A distinct population-level inflection point emerged at $\mathrm{THI} \approx 76$, beyond which both median and mean yields declined continuously, accompanied by a downward shift and narrowing of the IQR.

Evaluating cumulative losses relative to the comfort benchmark (Fig.~\ref{fig:response}b) demonstrated that production penalties remained negligible throughout the alert and mild zones.
Substantive losses began accumulating only after crossing THI 76, reaching an average reduction of approximately 11--13\% as conditions approached the severe boundary.
Bins with insufficient sample support ($n < 100$, comprising $<0.02\%$ of records above THI 80.5) were excluded to prevent compositional artefacts from small local cohorts.

Pairwise comparisons across discrete stress zones (Fig.~\ref{fig:response}c and Table~\ref{tab:ttest}) further clarified the magnitude and practical relevance of these thermal transitions.
While differences between adjacent mild zones reached extreme statistical significance due to the massive statistical power of the dataset ($N > 200{,}000$ per zone), their standardised effect sizes were negligible: Comfort versus Alert yielded $\Delta = -0.26$~kg~animal$^{-1}$ d$^{-1}$ ($t_{\mathrm{stat}} = -10.58$, $p = 3.9 \times 10^{-26}$, Cohen's $d = -0.03$), and Alert versus Mild showed $\Delta = 0.26$~kg~animal$^{-1}$ d$^{-1}$ ($t_{\mathrm{stat}} = 10.39$, $p = 2.7 \times 10^{-25}$, Cohen's $d = 0.03$).
In sharp contrast, the transition from Mild to Moderate stress marked the primary biological inflection point, producing a substantial yield contraction of $\Delta = 1.01$~kg~animal$^{-1}$ d$^{-1}$ ($t_{\mathrm{stat}} = 20.98$, $p = 1.2 \times 10^{-97}$) and a standardised effect size of Cohen's $d = 0.12$---four-fold larger than those observed across cooler categories.
Although conventional statistical benchmarks classify $d \approx 0.12$ as a small effect, this metric reflects an individual daily reduction ($\approx 1.0$~kg~animal$^{-1}$ d$^{-1}$) standardised against the large gross population variance in test-day yield ($\mathrm{SD} = 8.82$~kg~animal$^{-1}$ d$^{-1}$).
In commercial dairy farming, an individual deficit of this magnitude aggregates rapidly over time and across herds (amounting to $\approx 500$~kg of uncollected milk daily for a 500-animal herd), carrying substantial biophysical and economic importance that is further amplified in high-producing animals.
Finally, daily yield dropped an additional $0.37$~kg~animal$^{-1}$ d$^{-1}$ between Moderate and Severe zones ($d = 0.04$), but this contrast was not statistically significant ($t_{\mathrm{stat}} = 1.04$, $p = 0.30$), reflecting the scarcity of extreme weather events in commercial test-day records ($N = 584$).

\begin{table}[htbp]
\centering
\caption{Comparison of daily milk yield between adjacent thermal zones (independent-samples Student's $t$-test). $\Delta$ is the mean of the cooler zone minus the mean of the warmer zone.}
\label{tab:ttest}
\begin{tabular}{lrrrr}
\hline
\textbf{Comparison} & \textbf{$\Delta$ (kg~animal$^{-1}$ d$^{-1}$)} & \textbf{$t_{\mathrm{stat}}$} & \textbf{$p$} & \textbf{Cohen's $d$} \\
\hline
Comfort vs. Alert   & $-0.26$ & $-10.58$ & $3.9 \times 10^{-26}$ & $-0.03$ \\
Alert vs. Mild      & $0.26$  & $10.39$  & $2.7 \times 10^{-25}$ & $0.03$ \\
Mild vs. Moderate   & $1.01$  & $20.98$  & $1.2 \times 10^{-97}$ & $0.12$ \\
Moderate vs. Severe & $0.37$  & $1.04$   & $0.30$                & $0.04$ \\
\hline
\end{tabular}
\end{table}

\subsection{Aggregate correlations}
At the unadjusted record level, bivariate correlation between daily THI and milk yield was virtually negligible ($\rho = -0.011$, $r = -0.009$), whereas the negative association between DIM and production was more than an order of magnitude stronger ($\rho = -0.166$). This contrast underscores that individual stage of lactation and herd management account for vastly greater gross variance in test-day output than ambient weather fluctuations, highlighting the necessity of multi-variable panel models to isolate true thermal sensitivities.

\subsection{Lagged and cumulative effects of THI}
\begin{figure}[htbp]
    \centering
    \includegraphics[width=\linewidth]{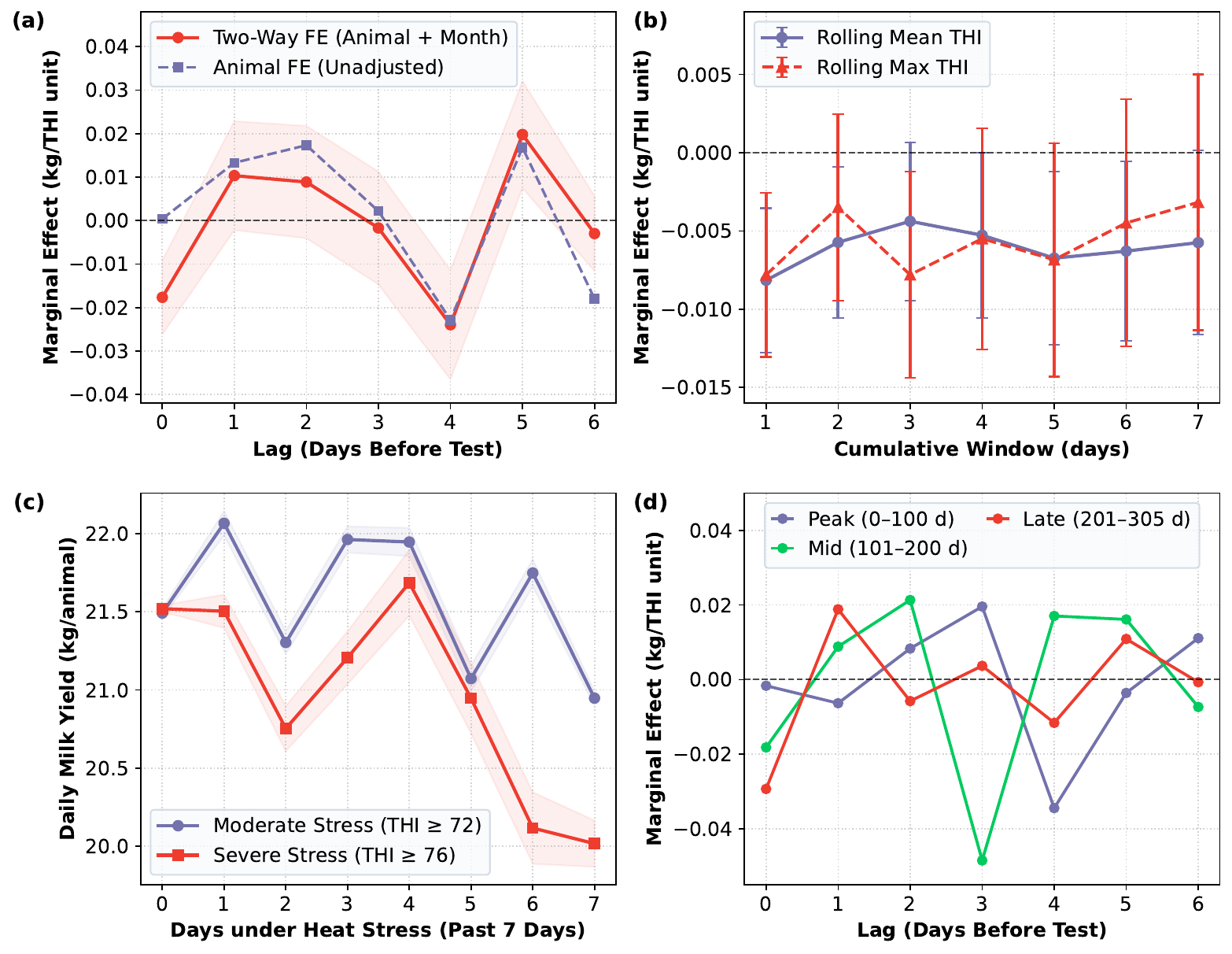}
    \caption{Lagged and cumulative associations between THI and daily milk yield. (a) Distributed lag response profile ($\beta_k \pm 95\%$ CI, kg~animal$^{-1}$ d$^{-1}$~unit$^{-1}$) for daily lags $t_0$ to $t_{-6}$, comparing Two-Way Fixed Effects (TWFE, animal and calendar month) with Animal Fixed Effects (FE) unadjusted for month. (b) Marginal effects ($\beta \pm 95\%$ CI) of single cumulative rolling windows (1 to 7 days) for rolling mean and rolling maximum THI under TWFE. (c) Observed daily milk yield (mean $\pm 95\%$ CI) as a function of the number of consecutive days under moderate ($\mathrm{THI} \ge 72$) and severe ($\mathrm{THI} \ge 76$) stress during the preceding seven days. (d) Lactation-stratified distributed lag response profiles for peak (0--100~d), mid (101--200~d) and late (201--305~d) lactation under TWFE.}
    \label{fig:lag_cumulative}
\end{figure}

The temporal dynamics of thermal exposure are illustrated in Fig.~\ref{fig:lag_cumulative}. In the TWFE DLM (Eq.~(\ref{eq:dlm}); Fig.~\ref{fig:lag_cumulative}a), controlling simultaneously for animal identity and calendar month revealed a distinct bimodal response profile. Heat exposure triggered an immediate suppression of milk yield on the test day itself ($t_0$), followed by an intermediate neutral period and a second, deeper negative trough four days later ($t_{-4}$). This delayed trough reflects the physiological latency between initial thermal strain, depressed dry matter intake, and the subsequent reduction in mammary secretory activity. Crucially, failing to account for seasonal calendar months resulted in spurious positive coefficients at intermediate lags, demonstrating that seasonal management cycles severely confound unadjusted estimates.

Evaluating cumulative exposure through multi-day rolling windows (Fig.~\ref{fig:lag_cumulative}b) confirmed consistent negative impacts across both mean and maximum daily THI, with maximum daytime heat driving pronounced acute losses. Furthermore, multi-day persistence of heat stress produced non-linear production penalties (Fig.~\ref{fig:lag_cumulative}c): while consecutive days under mild conditions caused modest oscillations, sustained exposure to severe conditions ($\mathrm{THI} \ge 76$) for five or more continuous days resulted in substantial and statistically significant yield losses.

Stratifying the lag profile by lactation stage (Fig.~\ref{fig:lag_cumulative}d) revealed clear biological contrasts in timing. High-producing animals in peak lactation (0--100~d) showed strong resilience to immediate same-day shocks but suffered their largest production drop four days after exposure. In contrast, animals in mid- and late lactation exhibited immediate yield reductions on the test day, with cumulative weekly losses persisting across the entire lag window.

\subsection{Diurnal thermal amplitude, nighttime cooling, and multi-night tropical persistence}
\begin{figure}[htbp]
    \centering
    \includegraphics[width=\linewidth]{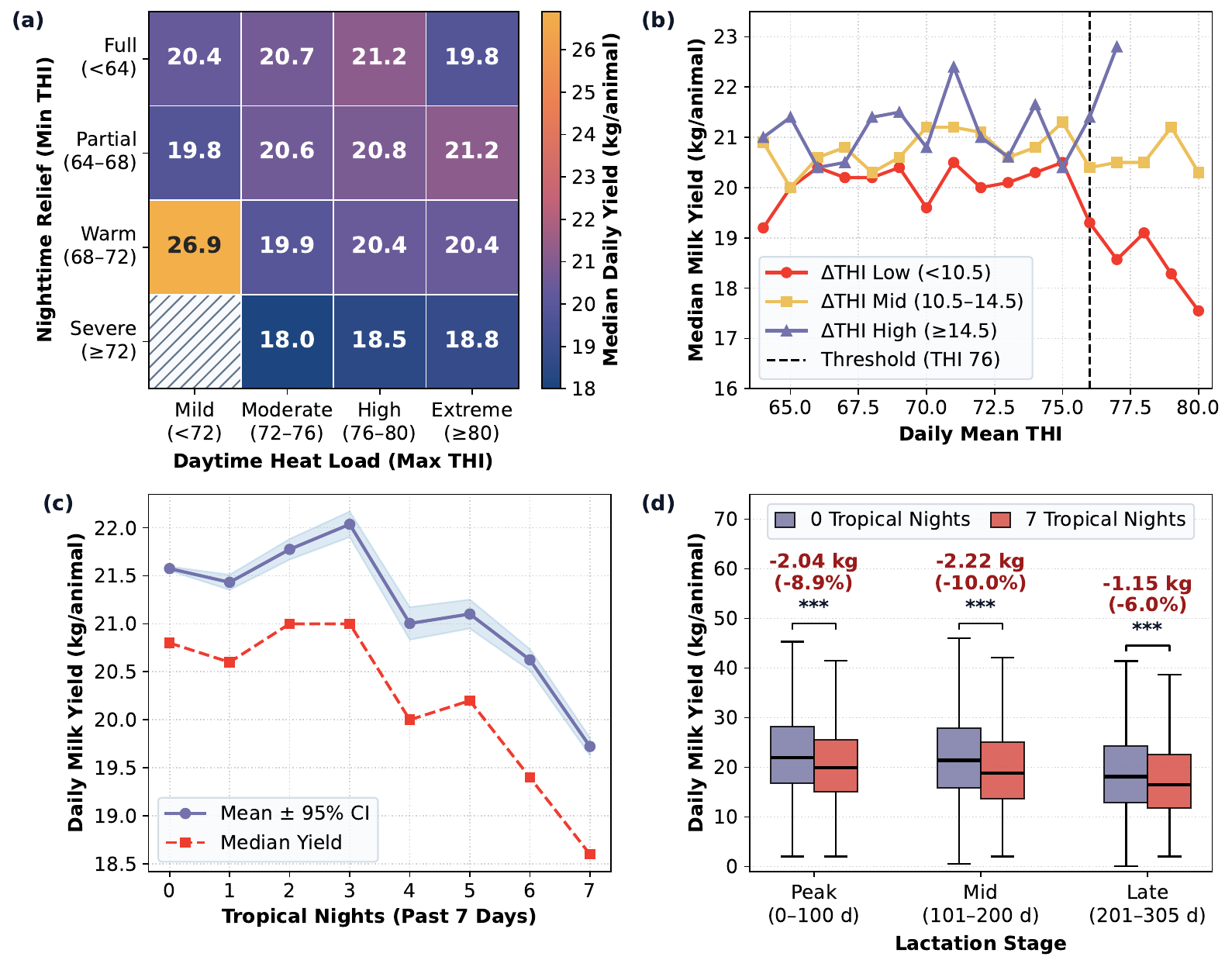}
    \caption{Nighttime thermal relief, diurnal thermal amplitude ($\Delta\mathrm{THI}$), and multi-night tropical persistence. (a) Median daily milk yield (kg~animal$^{-1}$ d$^{-1}$) across a two-dimensional interaction grid of daytime heat ($\mathrm{THI}_{\max}$) and nocturnal relief ($\mathrm{THI}_{\min}$). (b) Protective buffering of diurnal thermal amplitude (Low $<10.5$, Medium $10.5\text{--}14.5$, High $\ge 14.5$) across continuous daily mean THI (1.0-unit bins from 64 to 80), showing substantial yield recovery under heat stress when amplitude is high. (c) Cumulative effect of multi-night tropical persistence (0 to 7 tropical nights with $\mathrm{THI}_{\min} \ge 68$ in the preceding seven days) on daily milk yield (mean $\pm 95\%$ CI and median). (d) Lactation-stratified response comparing zero versus seven consecutive tropical nights using boxplots (outliers omitted, central line indicating median and box bounds indicating IQR). Pairwise brackets display statistical significance ($^{***} p \le 0.001$, two-sided Student's $t$-test) with labels displaying the mean yield difference and relative percentage.}
    \label{fig:nighttime_relief}
\end{figure}

The interplay between daytime peak heat, nocturnal cooling relief, and diurnal thermal amplitude is presented in Fig.~\ref{fig:nighttime_relief}. Approximately one-fifth (21.3\%) of all test-day records coincided with tropical nights ($\mathrm{THI}_{\min} \ge 68$), where nocturnal conditions completely precluded heat dissipation.

The two-dimensional interaction between daytime maximum heat and nocturnal minimum THI demonstrated that nighttime relief is a primary determinant of dairy resilience (Fig.~\ref{fig:nighttime_relief}a). Under identical high daytime heat, animals that experienced cool nocturnal recovery maintained baseline productivity, whereas animals subjected to tropical nights suffered an additional yield loss of over 2~kg~animal$^{-1}$ d$^{-1}$ (2.2~kg~animal$^{-1}$ d$^{-1}$ in median yield and 2.37~kg~animal$^{-1}$ d$^{-1}$ in mean yield). Even under extreme daytime heat ($\mathrm{THI}_{\max} > 80$), moderate nighttime cooling substantially buffered milk output compared to suffocating nights.

Consistent with this mechanism, diurnal thermal amplitude exerted a powerful protective effect during warm periods (Fig.~\ref{fig:nighttime_relief}b). While thermal amplitude had minimal bearing on production under comfortable weather, wide diurnal swings provided substantial yield buffering under moderate daytime heat, elevating daily output by $+1.60$~kg~animal$^{-1}$ d$^{-1}$ compared to environments with narrow amplitude. In the TWFE econometric specification (Eq.~(\ref{eq:night})), the penalty for tropical nights remained highly significant after controlling for peak daytime heat, animal fixed effects, and seasonal trends.

The cumulative burden of consecutive tropical nights generated a steep and monotonic yield depression (Fig.~\ref{fig:nighttime_relief}c). Daily production declined steadily with each additional tropical night, reaching an unadjusted loss of 1.85~kg~animal$^{-1}$ d$^{-1}$ ($-8.6\%$) after seven consecutive nights without nocturnal relief. Lactation-stratified comparisons (Fig.~\ref{fig:nighttime_relief}d) confirmed that this cumulative nocturnal burden fell most heavily on peak- and mid-lactation animals, which suffered yield reductions roughly double those observed in late-lactation herdmates.

\subsection{Production-level heterogeneity and empirical threshold shifts}
\begin{figure}[htbp]
    \centering
    \includegraphics[width=\linewidth]{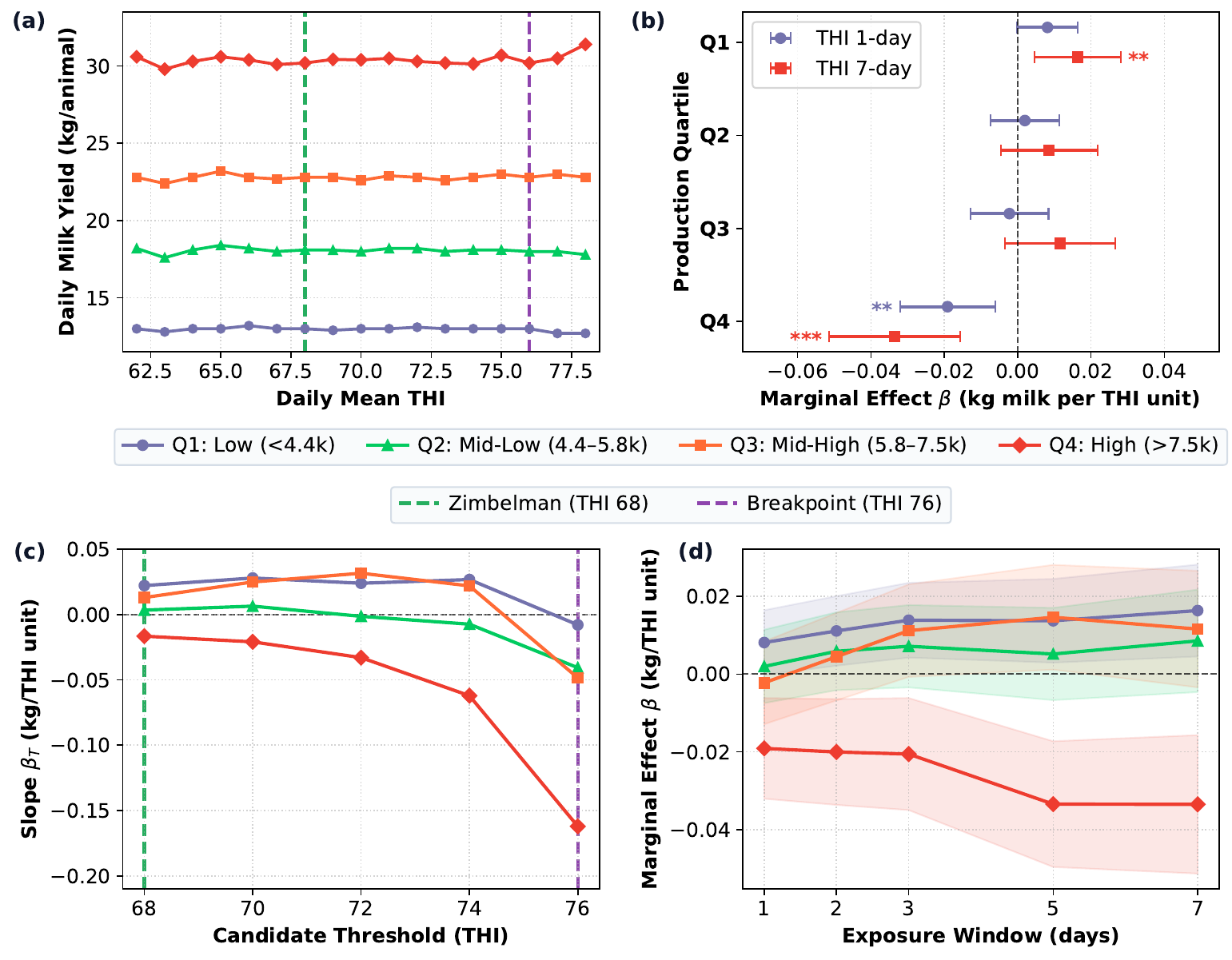}
    \caption{Heterogeneity in heat stress response by mature-equivalent 305-day milk production potential ($\mathrm{PROD\_305}$). (a) Observed daily milk yield (median $\pm$ point marker) across continuous daily mean THI for production quartiles ($Q_1$ Low $<4,389$~kg, $Q_2$ Mid-Low $4,389\text{--}5,847$~kg, $Q_3$ Mid-High $5,847\text{--}7,494$~kg, $Q_4$ High $>7,494$~kg), referencing the \citet{zimbelman2009} threshold ($\mathrm{THI} = 68$) and the aggregate inflection point ($\mathrm{THI} = 76$). (b) Econometric TWFE marginal effects ($\beta \pm 95\%$ CI) of immediate daily THI (1-day) and cumulative weekly THI (7-day) across quartiles, with asterisks indicating statistical significance ($^{***} p \le 0.001$, $^{**} p \le 0.01$, $^{*} p \le 0.05$). (c) Piecewise linear threshold sensitivity profile ($\beta_T$, kg~animal$^{-1}$ d$^{-1}$~unit$^{-1}$ above candidate threshold $T \in \{68, 70, 72, 74, 76\}$), highlighting the pronounced collapse for high-yielding animals ($Q_4$) under elevated thermal loads. (d) Cumulative rolling exposure dynamics ($\beta \pm 95\%$ CI) across expanding rolling windows (1, 2, 3, 5, and 7 days) under TWFE, showing progressive thermal exhaustion in $Q_4$.}
    \label{fig:production_heterogeneity}
\end{figure}

To evaluate whether production level affects heat stress sensitivity, animals were stratified into quartiles of 305-day milk yield ($Q_1$ to $Q_4$, Fig.~\ref{fig:production_heterogeneity}). The empirical dose--response curves (Fig.~\ref{fig:production_heterogeneity}a) revealed a striking divergence across production tiers: while low- and moderate-yielding animals ($Q_1$ to $Q_3$) maintained stable yields up to the aggregate population threshold of THI 76, animals in the top production quartile ($Q_4$) exhibited earlier instability and a steep downward trajectory under elevated temperatures.

Econometric panel models (Eq.~(\ref{eq:twfe})) controlling for individual animal fixed effects and lactation curves confirmed that thermal sensitivity scaled monotonically with productivity (Fig.~\ref{fig:production_heterogeneity}b). Under acute single-day exposure, low-producing animals showed no adverse response, whereas high-producing animals incurred significant production drops. Under cumulative weekly exposure, this disparity widened dramatically: the marginal penalty for top-producing animals nearly doubled ($\beta_{7\mathrm{d}} = -0.0337$~kg~animal$^{-1}$ d$^{-1}$~unit$^{-1}$, $p = 0.00021$), reflecting progressive physiological exhaustion under sustained heat load.

Piecewise linear splines across candidate thresholds (Eq.~(\ref{eq:spline}); Fig.~\ref{fig:production_heterogeneity}c) provided conclusive evidence of an empirical threshold shift. Low-producing animals maintained neutral or slightly positive trajectories across all candidate temperatures ($\beta_{76} = -0.0076$, $p = 0.81$). In stark contrast, high-producing animals began experiencing production losses at the classical temperate threshold of THI 68 ($\beta_{68} = -0.0167$, $p = 0.082$), with losses accelerating above THI 72 ($\beta_{72} = -0.0331$, $p = 0.048$) and plunging precipitously beyond THI 76 ($\beta_{76} = -0.1622$~kg~animal$^{-1}$ d$^{-1}$~unit$^{-1}$, $p = 0.0062$). Furthermore, multi-day rolling windows (Fig.~\ref{fig:production_heterogeneity}d) demonstrated that cumulative heat exposure worsened production losses exclusively in high-producing animals, establishing that baseline productivity is a primary driver of bioclimatic vulnerability in commercial tropical herds.

\subsection{Breed composition, heterosis, and thermal resilience across dairy crossbreds}
\begin{figure}[htbp]
    \centering
    \includegraphics[width=\linewidth]{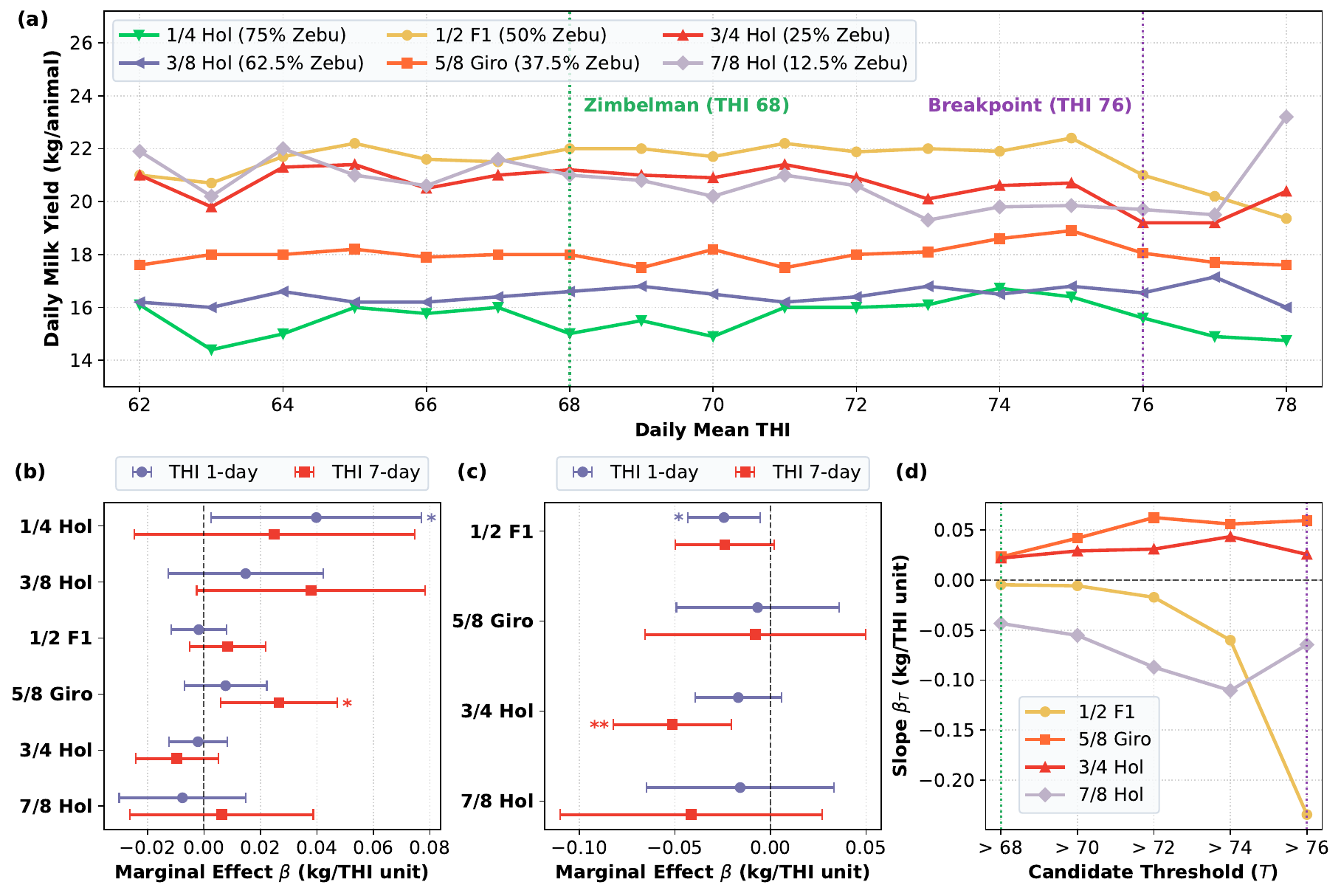}
    \caption{Heterogeneity in heat stress response by breed composition and heterosis in Holstein $\times$ Gyr crossbreds. (a) Panoramic continuous dose--response curve of observed daily milk yield (median $\pm$ point marker) across daily mean THI (62--78) for six genetic groups ($1/4$~Hol to $7/8$~Hol), referencing \citet{zimbelman2009} ($\mathrm{THI} = 68$) and the tropical breakpoint ($\mathrm{THI} = 76$). (b) Econometric TWFE marginal effects ($\beta \pm 95\%$ CI) of immediate daily THI (1-day) and cumulative weekly THI (7-day) across the full population. (c) High-yield decoupling stress test within the top production quartile ($Q_4$: $\mathrm{PROD\_305} \ge 7,494$~kg, all animals yielding $\approx 31$~kg~animal$^{-1}$ d$^{-1}$), demonstrating thermal buffering in $1/2$~F1 and $5/8$~Girolando versus severe cumulative breakdown in $3/4$~Hol ($\beta_{7\mathrm{d}} = -0.051^{**}$) and $7/8$~Hol. (d) Piecewise linear threshold trajectories ($\beta_T$, kg~animal$^{-1}$ d$^{-1}$~unit$^{-1}$ above candidate thresholds $T \in \{68, 70, 72, 74, 76\}$), highlighting contrasting biological breakpoints: $7/8$~Hol collapses early at $\mathrm{THI} = 68$, whereas $1/2$~F1 sustains production until $\mathrm{THI} = 76$.}
    \label{fig:breed_heterogeneity}
\end{figure}

Evaluating thermal sensitivity across Holstein $\times$ Gyr crossbred groups ($1/4$~Hol to $7/8$~Hol; Fig.~\ref{fig:breed_heterogeneity}) demonstrated that tropical adaptation and heterosis substantially modulate bioclimatic vulnerability. In population dose--response curves (Fig.~\ref{fig:breed_heterogeneity}a), intermediate crossbreds ($1/2$~F1 and $5/8$~Girolando) maintained level or slightly rising median yields across the entire range up to THI 76, confirming the broad thermal buffer conferred by Zebu genetics. Across the full population, econometric panel models (Fig.~\ref{fig:breed_heterogeneity}b) showed neutral responses to thermal shocks in Zebu-rich groups, whereas Holstein-dominated groups exhibited negative marginal responses that intensified under weekly cumulative exposure.

Because breed composition is naturally correlated with baseline yield, a decoupling stress test was conducted by restricting the analysis to animals in the highest production quartile, where animals operated at identical high daily yields of approximately 31~kg~animal$^{-1}$ d$^{-1}$ (Fig.~\ref{fig:breed_heterogeneity}c). Under sustained weekly heat, intermediate crossbreds ($5/8$~Girolando and $1/2$~F1) showed remarkable resilience, sustaining production with negligible or modest losses ($\beta_{7\mathrm{d}} = -0.008$) despite their high milk output. In sharp contrast, high-producing $3/4$~Hol and $7/8$~Hol animals suffered severe, statistically significant declines ($\beta_{7\mathrm{d}} = -0.051$, $p = 0.0011$ for $3/4$~Hol), confirming that retained Zebu heterosis provides genuine physiological heat tolerance independent of production level.

Finally, piecewise threshold splines (Fig.~\ref{fig:breed_heterogeneity}d) revealed contrasting biological tipping points across genetic groups. Animals with $7/8$~Hol fraction broke immediately at the temperate threshold of THI 68 ($\beta_{68} = -0.043$), with losses accelerating under higher temperatures. In striking contrast, $1/2$~F1 crossbreds remained fully buffered up to THI 76, declining only when ambient conditions exceeded this extreme tropical threshold. These results demonstrate that the elevated aggregate threshold of the national herd is underpinned by the biophysical resilience of crossbred genetics.

%% ===================================================================
\section{Discussion}\label{sec:discussion}
%% ===================================================================

\subsection{An empirical inflection at THI $\approx$ 76, production-level shifts, and crossbred heterosis}
The most consistent finding across the continuous and aggregate analyses is that population median daily yield remained essentially unchanged up to a daily mean THI of about 76 and declined thereafter. This value is higher than the threshold of 68 proposed for high-producing animals under controlled conditions \citep{zimbelman2009} and higher than the classical threshold of 72 \citep{armstrong1994, ravagnolo2000}. Several factors contribute to this difference. First, the daily \emph{mean} THI averages cooler nighttime conditions with daytime peaks, so a daily mean of 76 implies considerably higher maximum values during the day; thresholds defined on maximum or hourly THI are therefore not directly comparable \citep{bohmanova2007}. Second, herds in hot regions may have adopted shading, ventilation or other mitigation measures, or may contain crossbred animals with greater heat tolerance, which raises the apparent threshold in field data \citep{schutz2011, almeida2013}.

Crucially, the empirical stratifications by individual production potential (Fig.~\ref{fig:production_heterogeneity}) and breed composition (Fig.~\ref{fig:breed_heterogeneity}) resolve the longstanding question of why national tropical thresholds appear higher than North American experimental values. Endogenous metabolic heat generation ($H_{\mathrm{met}}$) is directly proportional to dry matter intake and milk synthesis, with high-yielding dairy animals generating up to double the heat load of average herdmates \citep{kadzere2002, west2003, collier2006, silva2002}. In the present sample, the average animal produced 21.5~kg~animal$^{-1}$ d$^{-1}$, and animals in the lowest production quartile ($Q_1$, 13.7~kg~animal$^{-1}$ d$^{-1}$) exhibited virtually zero negative sensitivity even at $\mathrm{THI} > 76$ ($\beta_{76} = -0.0076$, $p = 0.81$, Fig.~\ref{fig:production_heterogeneity}c) because their modest metabolic heat load is easily shed under tropical conditions.

In sharp contrast, when evaluating high-producing animals ($Q_4$, mean yield 30.8~kg~animal$^{-1}$ d$^{-1}$), the thermal comfort threshold dropped to $\mathrm{THI} \approx 68\text{--}72$ ($\beta_{68} = -0.0167$, $p = 0.082$; $\beta_{72} = -0.0331$, $p = 0.048$), with acute losses reaching $-0.1622$~kg~animal$^{-1}$ d$^{-1}$~unit$^{-1}$ ($p = 0.0062$) above $\mathrm{THI} = 76$. Furthermore, cumulative exposure over 5 to 7 days doubled the marginal loss in $Q_4$ to $-0.0337$~kg~animal$^{-1}$ d$^{-1}$~unit$^{-1}$ ($p = 0.00021$, Fig.~\ref{fig:production_heterogeneity}d). These empirical findings in commercial herds corroborate recent high-resolution bioclimatic modelling indicating that projected milk production declines under severe heat stress escalate disproportionately in high-producing Brazilian dairy cattle \citep{FERREIRA2027105027}.

Beyond baseline production level, the Brazilian dairy herd is uniquely underpinned by \emph{Bos indicus} (Gyr) crossbreeding, which confers superior cutaneous evaporative capacity, lower basal metabolic rates, and pronounced heterosis \citep{titto2011, costa2020, santana2017}. As demonstrated in Fig.~\ref{fig:breed_heterogeneity}c, even when high-producing animals operate at identical peak levels ($\approx 31$~kg~animal$^{-1}$ d$^{-1}$), $1/2$~F1 and $5/8$~Girolando crossbreds withstand cumulative weekly thermal stress without significant loss ($\beta_{7\mathrm{d}} = -0.008$), whereas $3/4$~Hol animals suffer severe metabolic breakdown ($\beta_{7\mathrm{d}} = -0.051$, $p = 0.0011$). In addition, $7/8$~Hol animals break already at the \citet{zimbelman2009} threshold of $\mathrm{THI} = 68$ ($\beta_{68} = -0.043$, Fig.~\ref{fig:breed_heterogeneity}d), while $1/2$~F1 animals resist thermal decline until $\mathrm{THI} = 76$. These results demonstrate that the aggregate tropical breakpoint of $\mathrm{THI} \approx 76$ is jointly driven by moderate herd average yields and by the biological thermal buffer of Zebu crossbreeding and heterosis \citep{silva2021}. High-grade Holsteins in warm climates operate under the exact same biological threshold ($\mathrm{THI} \approx 68$) established in temperate confinement, demonstrating that genetic selection for high Holstein fraction without comprehensive environmental cooling creates acute bioclimatic vulnerability.

The magnitude of the decline above THI 76 (about 1.7~kg~animal$^{-1}$ d$^{-1}$, or 8\%, between 76 and 80) and the 15--17\% reduction in median yield under Severe conditions in the first half of lactation are consistent in direction with the production losses reported for heat-stressed animals in experimental and field studies \citep{west2003, rhoads2009, bernabucci2014}. Nevertheless, point estimates in the Severe zone ($\mathrm{THI} > 80$, $N = 584$, $0.07\%$ of records) and above THI 80.5 should be interpreted with caution due to restricted sample support in the upper thermal tail.

\subsection{Heat stress and lactation stage}
The decline under Moderate and Severe conditions was concentrated in peak and mid-lactation stages, when milk synthesis, feed intake and metabolic heat production are highest, and it was absent in late lactation. This pattern agrees with the physiological expectation that animals at high production levels are more susceptible to heat stress \citep{kadzere2002, collier2006}. The higher median yield observed in late-lactation records under Severe conditions (19.7~kg~animal$^{-1}$ d$^{-1}$ vs. 17.4~kg~animal$^{-1}$ d$^{-1}$ in Moderate stress) does not represent biological thermotolerance; rather, it reflects compositional selection bias within a small subsample ($N = 213$) concentrated in a few intensive herds and years where high-performing animals were selectively retained in late lactation, illustrating the limits of unadjusted cross-sectional comparisons in sparse cells.

\subsection{Confounding in observational field data}
Several results indicate that the raw association between THI and yield is strongly confounded. The aggregate correlation between THI and yield was close to zero, the Alert zone showed slightly higher yield than the Comfort zone, and monthly mean yield peaked in September--October rather than in the coolest months. These patterns are consistent with seasonal calving: animals calving in autumn and winter reach peak lactation in the cool season, whereas animals recorded in summer are, on average, at a more advanced stage of lactation, as reflected in the seasonal variation of mean DIM. The steady long-term increase in yield (0.56~kg~animal$^{-1}$ d$^{-1}$ per year) further implies that unadjusted comparisons between thermal zones are affected by when the records were collected; this is particularly relevant for the Severe zone, almost half of whose records were collected in 2019 and 2023. Finally, the spatial analysis shows that heat-stress exposure is strongly tied to geography, so that differences between zones also partly reflect differences between regions, production systems and herds, in line with broad geographical shifts in thermal bioclimates and cattle heat-stress exposure across Brazil \citep{Ferreira:2026aa}.

The comparison of zones also illustrates the distinction between statistical and practical significance in large datasets. With hundreds of thousands of records, differences of a quarter of a kilogram per day are highly significant, but the corresponding effect sizes are negligible. Conversely, the Moderate--Severe contrast, which involved a difference of similar magnitude, was not significant ($p = 0.30$) because of the small number of Severe records. Crucially, these descriptive confounders motivate the TWFE panel specifications: by demeaning within individual animals and including calendar-month or month-by-year period fixed effects ($\gamma_{\mathrm{period}(t)}$), the econometric models effectively absorb both macro-seasonal reproductive cycles and the 24-year secular genetic and technological trend.

\subsection{Temporal dynamics: physiological latency, accumulation, and the scale of marginal effects}
The distributed lag and cumulative analyses provide two major insights into how heat stress operates over time in commercial dairy herds.

First, the bimodal lag structure observed in Fig.~\ref{fig:lag_cumulative}a---characterised by an acute drop on the test day ($t_0$) and a delayed trough at lag~4 ($t_{-4}$)---is consistent with ruminant physiology, even though test-day records are sampled monthly and therefore relate point-in-time production to preceding daily weather rather than continuous daily milk weights. In the short term, thermal discomfort directly suppresses dry matter intake and triggers peripheral vasodilation on the day of heat exposure \citep{west2003, collier2006}. However, the broader metabolic and endocrine cascade---including down-regulation of thyroid hormones, suppression of the somatotropic and insulin-like growth factor 1 (IGF-1) axis, increased gut permeability and altered ruminal volatile fatty acid production---requires 72 to 96 hours to fully manifest as reduced secretory activity in mammary epithelial cells \citep{rhoads2009, bernabucci2014, tao2018}. The lactation-stratified profiles (Fig.~\ref{fig:lag_cumulative}d) further clarify this mechanism: animals in peak lactation possess high metabolic drive and initially mobilise adipose tissue reserves to cushion the immediate shock, postponing the yield decline to lag~4; in contrast, animals in mid and late lactation reduce milk synthesis immediately upon thermal exposure.

Second, the numerical magnitude of linear marginal effects ($\beta \approx -0.008$ to $-0.034$~kg~animal$^{-1}$ d$^{-1}$~unit$^{-1}$) might appear modest at first glance, representing only 8 to 34~g of milk per THI unit. However, this appearance is driven by methodological and biological factors rather than an absence of real impact:
\begin{itemize}
    \item \emph{Scale of environmental variation:} Environmental THI does not fluctuate by single units in commercial farming. A standard summer heatwave shifts municipal conditions from thermoneutral comfort ($\mathrm{THI} \approx 64$) to severe stress ($\mathrm{THI} \ge 78$), an amplitude of 14--16 THI units. Multiplying the linear lag slope by this realistic excursion yields losses of 0.35 to 0.55~kg~animal$^{-1}$ d$^{-1}$ from linear thermal load alone.
    \item \emph{Linear attenuation across thermoneutral ranges:} Fitting a single linear slope across the entire THI spectrum averages the large thermoneutral region ($\mathrm{THI} < 68$), where thermal variation has zero or mildly positive associations with yield, with the hot range. This pulls the overall slope towards zero. In contrast, non-linear threshold persistence (Fig.~\ref{fig:lag_cumulative}c) and multi-night persistence (Fig.~\ref{fig:nighttime_relief}c) reveal that prolonged heatwaves over 6--7 consecutive days lead to cumulative losses of 1.50 to 1.85~kg~animal$^{-1}$ d$^{-1}$ (7.0\% to 8.6\% reductions in output).
    \item \emph{Econometric absorption in Two-Way Fixed Effects:} By including calendar-month or month-by-year indicators alongside individual animal fixed effects, the panel specifications intentionally absorb both macro-seasonal climate contrasts (the baseline summer--winter difference) and the long-term secular productivity trend ($+0.56$~kg~animal$^{-1}$ d$^{-1}$ per year). Consequently, the estimated $\beta$ reflects only high-frequency, residual weather departures within a given month.
    \item \emph{Herd-level and cumulative economic significance:} In commercial dairy operations, modest individual daily losses aggregate rapidly. For a typical herd of 500 animals, an individual reduction of 0.50 to 1.0~kg~animal$^{-1}$ d$^{-1}$ represents 250 to 500~kg of uncollected milk daily, amounting to 7,500 to 15,000~kg of lost production over a single 30-day summer month.
\end{itemize}
Together, these findings underscore that heat stress management cannot rely solely on same-day thermometer readings: cooling protocols must be maintained across sustained multi-day heat episodes to prevent cumulative metabolic breakdown.

\subsection{The biophysical and management significance of nighttime thermal relief}
The empirical results demonstrate that nocturnal conditions and diurnal thermal amplitude play a decisive role in modulating heat stress in dairy cattle under tropical and subtropical field conditions. In dairy cattle biophysics, heat balance is governed by the equilibrium between metabolic heat generation and environmental heat exchange \citep{kadzere2002, collier2006}. During hot daylight hours, high ambient temperatures and solar radiation markedly restrict sensible heat dissipation (convection and radiation), forcing lactating animals to rely almost exclusively on evaporative cooling through sweating and panting \citep{west2003, polsky2017}. Because high relative humidity further impedes cutaneous evaporation, animals accumulate a substantial internal heat load during the day, causing core body temperature to rise above physiological thresholds ($\sim 39.0\text{--}39.5^\circ\text{C}$).

Under natural thermal cycles, the recovery of homeothermy depends on nocturnal ambient cooling \citep{scott1983, igono1992, honig2012}. When nighttime ambient temperatures decline sufficiently (typically below $\mathrm{THI} \approx 64\text{--}68$), the thermal gradient between the animal's body surface and the surrounding air re-establishes sensible heat dissipation, allowing accumulated thermal load to be shed, respiratory rates to normalise, and dry matter intake to recover during the cooler night hours \citep{flamenbaum2010}. However, when ambient conditions remain in the tropical night regime ($\mathrm{THI}_{\min} \ge 68$), this thermal gradient remains collapsed throughout the 24-hour cycle. As shown in Fig.~\ref{fig:nighttime_relief}a, animals facing high daytime heat combined with warm tropical nights experience an additional loss of 2.37~kg~animal$^{-1}$ d$^{-1}$ in mean yield (2.2~kg~animal$^{-1}$ d$^{-1}$ in median yield) compared to those enjoying cool nights. Entering a new daylight period with an unresolved heat debt compounds physiological exhaustion, accelerating endocrine down-regulation and secretory depression in mammary tissue \citep{rhoads2009, bernabucci2014}.

The protective buffering conferred by diurnal thermal amplitude ($\Delta\mathrm{THI}$, Fig.~\ref{fig:nighttime_relief}b) carries direct implications for bioclimatic zonation. In Brazilian regions characterised by high thermal amplitude (such as high-altitude tropical savannahs (Cerrado) in Minas Gerais and Goi\'as), daytime heat stress is partially compensated by brisk nocturnal cooling, providing an intrinsic physiological respite ($+1.60$~kg~animal$^{-1}$ d$^{-1}$ relative to low-amplitude environments under moderate heat). Conversely, in low-altitude or humid coastal regions where maritime humidity dampens thermal amplitude, animals face unremitting 24-hour heat loads, with multi-night tropical persistence (Fig.~\ref{fig:nighttime_relief}c--d) exerting a cumulative toll reaching $-1.85$~kg~animal$^{-1}$ d$^{-1}$ ($-8.6\%$) over seven consecutive tropical nights, concentrated particularly in high-producing peak- and mid-lactation animals.

\subsection{Systems-level implications for tropical dairy breeding, housing, and farm management}
From an agricultural systems perspective, the present findings demonstrate that climate resilience in tropical dairying emerges from the interaction among animal genetics, physiological state, farm cooling infrastructure, and regional microclimate, providing actionable criteria to support decision-making for breeders' associations, livestock housing engineers, farm advisors, and dairy producers.

First, the sharp contrast between crossbred genotypes (Fig.~\ref{fig:breed_heterogeneity}) demonstrates that genetic selection for high milk production in warm climates cannot proceed independently of environmental adaptation. While upgrading dairy cattle to high Holstein fractions ($\ge 3/4$ or $7/8$) increases individual genetic yield potential, it simultaneously lowers the thermal inflection threshold to temperate levels ($\mathrm{THI} \approx 68$) and dramatically amplifies vulnerability to sustained multi-day heatwaves ($\beta_{7\mathrm{d}} = -0.051$~kg~animal$^{-1}$ d$^{-1}$~unit$^{-1}$). In contrast, intermediate crossbreds ($1/2$~F1 and stabilised $5/8$~Girolando) maintain stable milk yields up to $\mathrm{THI} \approx 76$, even when yielding at peak levels identical to high-grade Holsteins ($\approx 31$~kg~animal$^{-1}$ d$^{-1}$). This highlights that retained Zebu heterosis and functional tropical adaptations (e.g., higher sweating rates, short sleek hair coats, and greater tissue heat tolerance) provide a vital biophysical buffer \citep{titto2011, almeida2013, costa2020}. Dairy breeding programmes in tropical regions should strategically capitalise on this resilience, avoiding indiscriminate upgrading toward high-fraction taurine cattle unless intensive environmental modification is economically feasible.

Second, while the municipal-level meteorological dataset does not directly observe farm-level housing infrastructure or internal airflow, the empirical demonstration that tropical nights ($\mathrm{THI}_{\min} \ge 68$) and consecutive warm nights severely depress milk yield (Fig.~\ref{fig:nighttime_relief}) has direct implications when interpreted alongside dairy housing engineering literature \citep{bewley2017, leso2020, damasceno2020, andrade2022}. In many commercial confinement facilities across Brazil (such as compost-bedded pack barns and free-stalls), automated ventilation and cooling systems are routinely programmed to reduce fan speeds or switch off after sunset to economise electricity. The bioclimatic results indicate that suspending cooling during tropical nights leaves daytime heat debt unresolved. Integrating these findings with established engineering guidelines suggests that commercial facilities in warm climates should maintain continuous nighttime ventilation---targeting literature-recommended airspeeds of $1.8\text{--}2.5$~m~s$^{-1}$ over resting areas and strategic late-evening cooling cycles when ambient humidity permits---and prioritise cooling resources for high-producing and early-to-mid lactation groups \citep{schutz2011, bewley2017, andrade2022}.

\subsection{Limitations and future work}
While this study leverages a large dataset of 788,528 test-day records, within-animal panel models, and circadian microclimate metrics, several limitations should be noted. First, meteorological exposure was assigned using municipal-level weather stations, which cannot capture farm-specific microenvironments, elevation differences within municipalities, housing typology (e.g., pasture, free-stall, or compost-bedded pack barns), or the presence and operational regimes of active on-farm cooling systems (such as fan airspeeds and sprinkler schedules). Second, because official milk-recording data are collected at monthly intervals, the distributed lag models evaluate how point-in-time test-day yields respond to preceding daily weather rather than tracking continuous day-to-day lactation kinetics of the same animal across consecutive days. Third, although the overall dataset is large, sample size in the extreme upper tail ($\mathrm{THI} > 80$, $N = 584$) remains limited, warranting caution when extrapolating beyond $\mathrm{THI} \approx 80$.

Future research should integrate municipal meteorological records with high-frequency on-farm internet-of-things (IoT) sensor networks and inline electronic milk meters to continuously quantify indoor housing microclimates (including internal THI and air velocity) alongside daily individual milk yield kinetics. Furthermore, combining precision physiological monitoring---such as rumination collars, activity tags, and reticulorumen bolus temperature sensors---with hourly heat-load duration curves will clarify the real-time dynamics of heat debt accumulation and nocturnal recovery. Finally, integrating genomic evaluations with historical weather data represents a promising avenue to identify specific quantitative trait loci and genomic regions underpinning nocturnal heat dissipation and crossbred thermotolerance in tropical dairy cattle.

%% ===================================================================
\section{Conclusions}\label{sec:conclusions}
%% ===================================================================
This study provides a national-scale assessment of heat stress exposure and its association with milk yield in Brazilian dairy herds, analysing 788,528 test-day records from 87,809 animals across 408 municipalities over a 24-year period (2001--2024). The empirical results resolve the apparent discrepancy between the high inflection point observed in aggregate field records ($\mathrm{THI} \approx 76$) and classical experimental guidelines ($\mathrm{THI} \approx 68$), showing that thermal sensitivity is governed by individual metabolic output and crossbred genetics. High-producing animals ($Q_4 > 7,494$~kg, mean 30.8~kg~animal$^{-1}$ d$^{-1}$) suffer significant production declines beginning at $\mathrm{THI} \approx 68\text{--}72$, with marginal losses reaching $-0.162$~kg~animal$^{-1}$ d$^{-1}$~unit$^{-1}$ above 76, whereas low-producing animals ($Q_1 < 4,389$~kg) exhibit no adverse response across the observed range. Furthermore, while high-grade Holsteins ($7/8$~Hol) experience yield declines starting at $\mathrm{THI} = 68$, intermediate crossbreds ($1/2$~F1 and $5/8$~Girolando) sustain milk synthesis up to $\mathrm{THI} \approx 76$ even when operating at identical peak yields ($\approx 31$~kg~animal$^{-1}$ d$^{-1}$), confirming the essential role of Zebu heterosis in tropical thermal buffering.

Crucially, the analysis demonstrates that daytime heat load is strongly modulated by delayed physiological responses and nocturnal recovery. Distributed lag models reveal a bimodal response combining an immediate test-day drop with a deeper delayed trough four days later, while the complete absence of nighttime thermal relief ($\mathrm{THI}_{\min} \ge 68$, observed in 21.3\% of records) exacerbates daytime yield losses by an additional $-2.37$~kg~animal$^{-1}$ d$^{-1}$ and imposes a cumulative penalty of $-1.85$~kg~animal$^{-1}$ d$^{-1}$ ($-8.6\%$) after seven consecutive tropical nights. Conversely, high diurnal thermal amplitude ($\Delta\mathrm{THI} \ge 14.5$) buffers yield by $+1.60$~kg~animal$^{-1}$ d$^{-1}$. Collectively, these findings indicate that effective heat stress mitigation in warm climates cannot rely solely on daytime temperature measurements; commercial herd management and facility engineering should implement multi-day cooling protocols, continuous nighttime ventilation, and balanced crossbreeding strategies to safeguard animal welfare and productive sustainability.

%% ===================================================================
\section*{CRediT authorship contribution statement}
\textbf{Leonardo Nascimento Ferreira}: Conceptualization, Methodology, Software, Formal analysis, Writing -- original draft, Writing -- review and editing. \textbf{Delane Ribas da Rosa}: Writing -- review and editing. \textbf{Renata Negri}: Data curation, Writing -- review and editing. \textbf{Marcos Vinicius Gualberto Barbosa da Silva}: Data curation, Writing -- review and editing. \textbf{Rafaella Resende Andrade}: Writing -- review and editing. \textbf{Nicole Costa Resende Ferreira}: Conceptualization, Methodology, Formal analysis, Validation, Writing -- review and editing.

\section*{Declaration of competing interest}
The authors declare that they have no known competing financial interests or personal relationships that could have appeared to influence the work reported in this paper.

\section*{Data availability}
The milk-recording data analysed in this study were provided by the Brazilian Association of Girolando Breeders under data sharing agreements for research purposes and are subject to commercial confidentiality. Meteorological data from INMET are publicly available.

\section*{Acknowledgements}
This work was supported by the National Council for Scientific and Technological Development (CNPq, Brazil) [grant numbers 446734/2024-1 and 448218/2025-9]; and the Research Support Foundation of the State of Goi\'as (FAPEG, Brazil).

%% ===================================================================
\bibliographystyle{elsarticle-harv}
\bibliography{references}

\end{document}